\documentclass[twoside,leqno,twocolumn]{article}

\usepackage[letterpaper]{geometry}

\usepackage{mystyle}
\usepackage{ltexpprt}
\usepackage{hyperref}
\usepackage{float}
\usepackage[title]{appendix}
\usepackage{cancel}
\usepackage{booktabs}
\usepackage{algorithm}
\usepackage{algpseudocode}

\begin{document}

\newcommand\relatedversion{}
\renewcommand\relatedversion{\thanks{The full version of the paper can be accessed at \protect\url{https://arxiv.org/abs/1902.09310}}} % Replace URL with link to full paper or comment out this line

\title{\Large Integrable Frame Fields using Odeco Tensors}
% \author{\textit{Authors hidden for review}}
\author{Mattéo Couplet\thanks{University of Louvain, Belgium.} \and Alexandre Chemin\footnotemark[1] \and Jean-François Remacle\footnotemark[1]}

\date{}

\maketitle

% Copyright Statement
% When submitting your final paper to a SIAM proceedings, it is requested that you include
% the appropriate copyright in the footer of the paper.  The copyright added should be
% consistent with the copyright selected on the copyright form submitted with the paper.
% Please note that "20XX" should be changed to the year of the meeting.

% Default Copyright Statement
\fancyfoot[R]{\scriptsize{Copyright \textcopyright\ 2023 by SIAM\\
Unauthorized reproduction of this article is prohibited}}

\begin{figure*}
    \includegraphics[angle=90,origin=c,width=.49\textwidth]{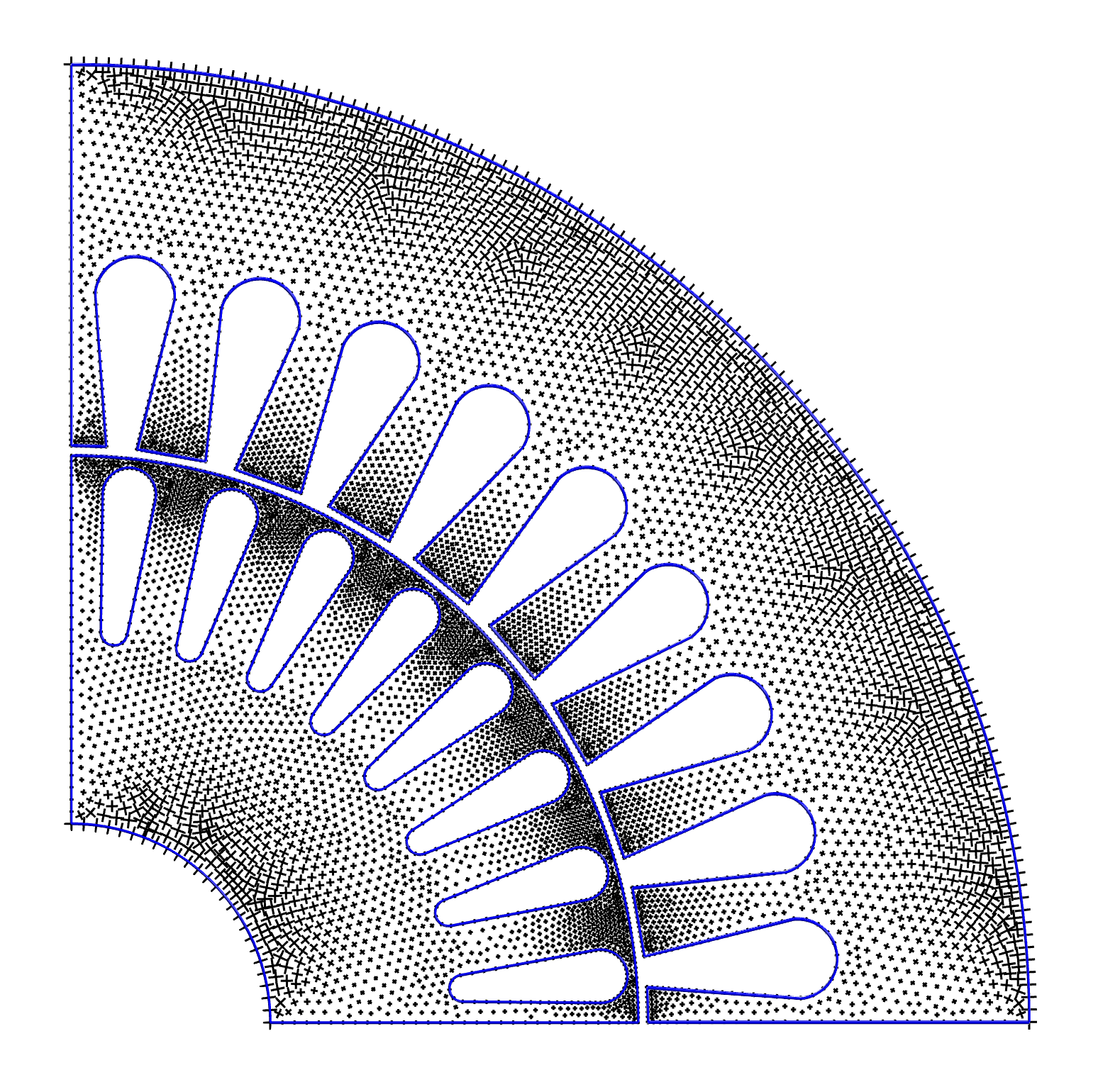}
    \includegraphics[width=.49\textwidth]{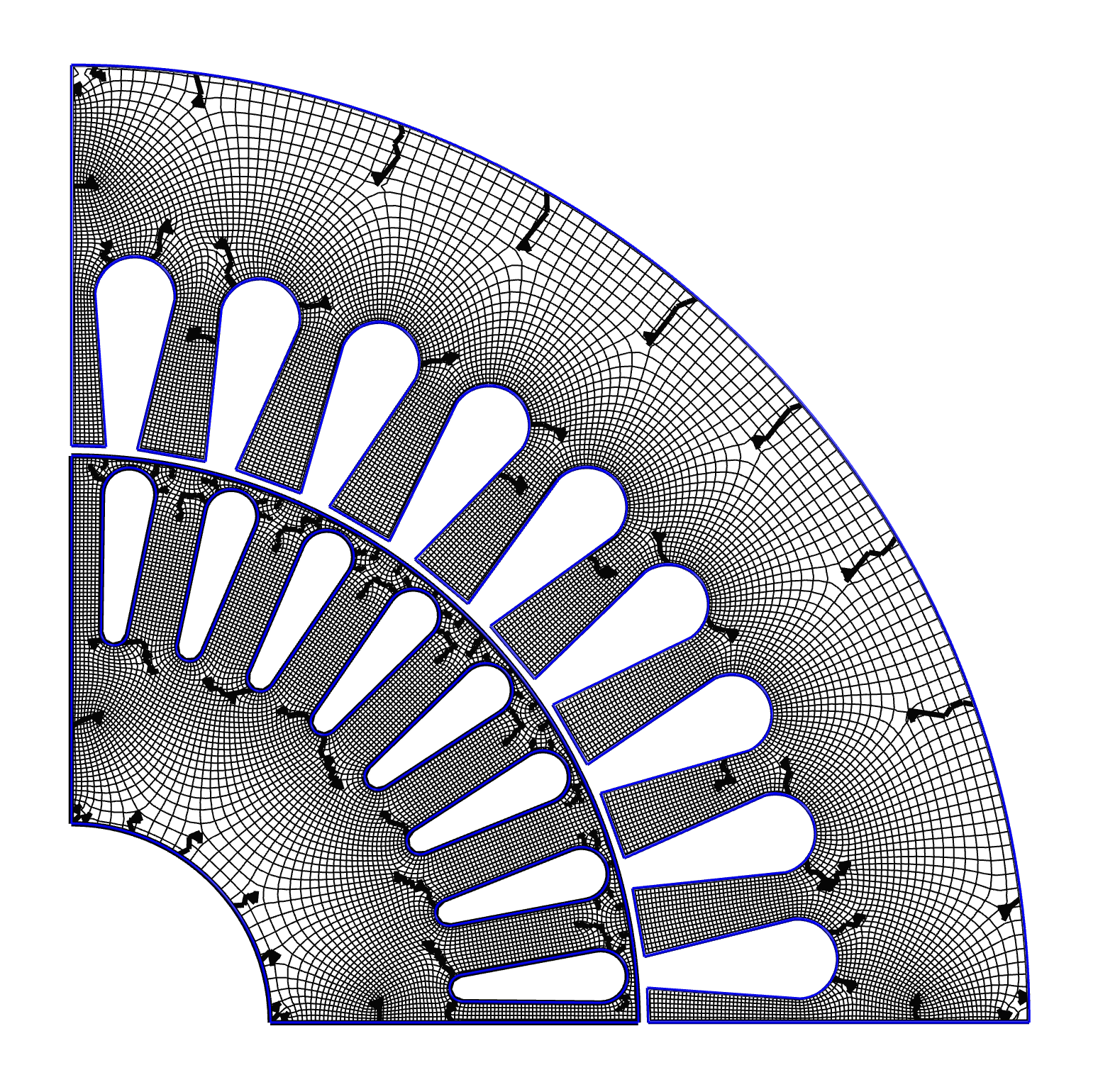}
    \caption{Overview of the approach on a machine composed of a rotor and stator.
    (Left) An integrable frame field is computed respecting user-prescribed size constraints:
    size of 1 on the inner and outer arcs, and size of \num{0.3} on the other boundaries.
    The frames indicate the local size and orientation of the quadrilateral mesh elements.
    (Right) A seamless parametrization is computed by integrating the frame field.
    Black triangles represent singularities and thick lines form the cut graph.}
    \label{fig:machine}
\end{figure*}

% Keywords: quadrilateral mesh, integrable frame field, integrability, cross field, parametrization, orthogonally decomposable tensor, odeco, spherical harmonics

\begin{abstract} \small\baselineskip=9pt 
We propose a method for computing integrable orthogonal frame fields on planar surfaces. Frames and their symmetries are implicitly represented using orthogonally decomposable (odeco) tensors. To formulate an integrability criterion, we express the frame field's Lie bracket solely in terms of the tensor representation; this is made possible by studying the sensitivity of the frame with respect to perturbations in the tensor. We construct an energy formulation that computes \update{smooth and integrable} frame fields, in both isotropic and anisotropic settings. The user can prescribe any size and orientation constraints in input, and the solver creates and places the singularities required to fit the constraints with the correct topology.
The computed frame field can be integrated to a seamless parametrization that is aligned with the frame field.
\end{abstract}

\section{Introduction} \label{sec:intro}

Meshes composed of quadrilaterals are known to offer superior performance
than their triangular counterpart when used as a support in numerical simulations,
however quad meshers still do not meet the same level of
robustness, flexibility and quality required
for industrial applications as triangular meshes.
In three dimensions, generating fully hexahedral meshes
is an even more difficult task,
and their industrial use is consequently very limited
despite a persistent demand from practitioners.

% Why is it hard?
Quadrilateral and hexahedral meshing are challenging because they couple
(a) a \emph{geometric} problem, minimizing the distortion of the elements,
and (b) a \emph{combinatorial} problem, achieving a conforming connectivity structure.
If one also desires a multi-block structure,
a layer of complexity is added, requiring (c) an adequately coarse block structure.
The last two decades have seen the emergence of \emph{field-based approaches},
which divide the problem in two main steps.
(1) The combinatorial constraints are ignored and a \emph{frame field}
is computed; a frame is a set of 2/3 directions
representing the orientation (and sometimes the size) of a quad/hex.
This frame field can be seen as a continuous extension of a mesh.
(2) A mesh is generated using guidance from the frame field.
This can be done through numerous approaches which we review below.

An issue shared by most existing works
is that no quad/hex mesh exactly follows the frame field computed a priori.
This is due to the frame field not being \emph{integrable};
we elaborate on this in~\autoref{sec:frame_fields}.
This limitation of frame field-based methods
means that the mesh deviates from the frame field
and the user must compromise on element quality, control over element size and orientation,
and control over the topology of the mesh.
A more fundamental issue arises when the computed frame field
has a global topology that is not meshable, making it unusable.
This happens in the 2D case when limit cycles appear;
in 3D the singular structure is very often invalid
due to the presence of non-meshable singular nodes.

Our contribution overcomes this limitation \update{in 2D}
and allows to generate frame fields \update{on planar surfaces} that are \update{\emph{integrable}
(\autoref{fig:machine}, left)};
during this process the user can prescribe any orientation or size constraints \update{along feature curves}.
Integrability guarantees that the frame field
can be integrated to a seamless parametrization
that is exactly aligned with the frame field and respects the prescribed constraints
\update{(\autoref{fig:machine}, right)}.
The parametrization can then be quantized to extract a quad mesh,
or the frame field can be directly used as a guide
to a frontal mesher; we illustrate the latter in our results.

Our approach relies on orthogonally decomposable (\emph{odeco}) tensors
which serve as an algebraic representation for the frames.
By studying the so-called \emph{eigenvalue sensitivity problem} for tensors,
we are able \update{to} formulate the problem completely in terms of this implicit algebraic representation.
This work is the first achieving integrability using odeco tensors.
As they can be extended to three dimensions,
this contribution paves the way for hex-meshable frame fields,
which are allegedly the key to the long-standing problem
of robustly generating optimal hexahedral meshes.

\subsection{Related work} \label{ssec:related_work}

\paragraph{Frame field design.}
4-direction fields, i.e., assignments of four directions
to every point of a surface, and their application to quad meshing
have been studied extensively in the computer graphics community;
paper \cite{vaxman16} provides a review.
Some important  works~\cite{ray08,crane10}
assume the field topology to be known in advance
and optimize for cross field smoothness.
If the topology is unknown,
it can be represented explicitly and optimized through integer variables
such as in~\cite{bommes09},
which often leads to mixed-integer formulations.
\update{Alternatively, the topology can also be implicitly encoded in the field representation;
our work fits in this category.
Earlier works in this line of research include~\cite{ray06,knoppel13}.
A key challenge is being able to represent singularities
while maintaining unit norm frames.
More recently, this problem has been re-framed
in the Ginzburg-Landau framework,
which replaces the ill-posed unit norm constraint by a penalty term
taken to the limit~\cite{beaufort17,viertel19}.}

\paragraph{Field-guided meshing.}
As an intermediate step before generating a quad mesh,
a global parametrization is often computed on the domain using the guidance from the frame field.
Among notable works we can cite~\cite{kalberer07,bommes09}
who integrate the frame field in a least-squares sense
to find the parametrization that best aligns to the frame field,
or~\cite{ray06}, who perform a curl-reduction procedure a posteriori.
Another approach, e.g., in~\cite{myles14},
is to trace out parametric lines from the frame field to form quadrilateral patches.
Getting a quad mesh from a global parametrization is typically
done using quantization methods, where a T-mesh is traced out
and its edges are given integer lengths; see, e.g.,~\cite{campen15,lyon21}.

Obtaining a quad mesh from a frame field can also be achieved
robustly using frontal method, by inserting points with guidance from the frame field.
Remeshing and smoothing is often necessary a posteriori to fix regions of bad quality;
see, e.g.,~\cite{reberol21}, which we use as a quad mesher in our pipeline.

In this work, the parametrization step is made trivial
as an integrable frame field is equivalent to a seamless parametrization.
Hence, integrating the frame field provides a parametrization
that is exactly aligned and matches the sizing of the frame field.
This confers the user a complete control over the sizing and orientation of the mesh,
a desirable property that cannot be achieved through conventional cross field guided meshing.

\paragraph{Frame representation.}
Computing frame fields with implicit topology
requires a representation that is invariant
to the ordering and symmetries of the frame vectors.
\update{The majority frame field-driven methods
use representations of \emph{unit} frames, or \emph{crosses},
as they only care about the orientation of the mesh;
a review can be found in~\cite{vaxman16}.
One of the most prominent representations
is the trigonometric pair $(\cos(4\theta), \sin(4\theta))$
(or, equivalently, the complex exponential $e^{i 4\theta}$)
which effectively encodes rotational symmetries
and is trivially normalized to unit norm~\cite{ray06,palacios07,kowalski13,knoppel13}.
However, this representation cannot encode sizes along with the directions.
On the other hand, PolyVectors, introduced by~\cite{diamanti14},
encode the frame vectors as the complex roots of a degree 4 polynomial.
They have the advantage that they can represent non-orthogonal frames (as done in~\cite{sageman-furnas19}),
but they offer no straightforward extension to volumetric frames.
For representing 3D frames, a spherical function is often used
that is maximal in the directions of the frame.
This function is encoded by its coefficients in the basis of spherical harmonics~\cite{huang11}.
As highlighted by~\cite{chemin19}, it can also be interpreted as a 4th-order tensor.
Recently, Palmer et al.~\cite{palmer20a} have proposed
an extension of these representations that encodes sizes along the directions,
relying on the so-called orthogonally decomposable (Odeco) tensors.
As they are the only known representation that can encode three-dimensional orthogonal and scaled frames,
we propose to use them in the 2D setting so that they can offer a 3D extension for later work.
}
% We focus on representations that encode \emph{sizes} for the frame vectors
% along with their directions.
% We rely on \update{the} two-dimensional version of the orthogonally decomposable (odeco) tensors,
% which were formalized by~\cite{robeva16}
% and later proposed as a frame representation by~\cite{palmer20a}.
% Odeco tensors are the only known representation that can encode
% three-dimensional orthogonal and scaled frames.
% Note that tensors can always be reinterpreted in terms of spherical harmonics
% through a change of basis.
% The main advantage of odeco tensors is that
% they belong to an algebraic variety that is fully described
% through a set of quadratic equations,
% and a robust projector onto the variety is available.
% This means that one does not need to include the frame vectors in the optimization variables.

\paragraph{Integrable frame fields.}
Motivated by the issue where the mesh is not aligned to the computed frame field,
a handful of works have focused on computing frame fields that are \emph{integrable},
i.e., the direction vectors are the gradients of a seamless parametrization.
This property holds if the two vector fields are \emph{curl-free}.
Diamanti et al.~\cite{diamanti15} achieve this using PolyVector fields.
Unlike our approach, the integrability condition cannot be expressed
in terms of the PolyVector coefficients,
and the optimization is done on the explicit direction vectors.
In~\cite{sageman-furnas19}, PolyVector fields are used to construct
so-called Chebyshev nets, i.e., quadrilateral meshes with a uniform size
but not necessarily orthogonal.
Our goal is slightly different (orthogonal quad meshing)
but we address the same challenges regarding control of size and orientation.
Like us, they optimize a frame field instead of a parametrization Jacobian,
and express the integrability through the Lie bracket.
The main advantage of odeco tensors compared to PolyVectors
is that the former offer a natural extension to 3D frame fields.

In~\cite{jezdimirovic22}, an integrable frame field is computed
given a user-imposed (or pre-computed) singularity configuration.
The frame fields are exactly integrable and the method can even remove
limit cycles that appear in most field-based approaches.
Having to specify singularities however is a significant limitation,
especially in the 3D case where almost all existing frame field solvers
produce invalid singular configurations.

% This requires special techniques to prevent inversions and preserve order of the direction vectors;
% this is not a concern in our framework.

% Most direction-field-based methods
% compute a seamless parametrization from the field
% by integrating it to a pair of scalar potentials.

% Our focus is on methods that do not assume the field topology to be known in advance

% \paragraph{Direction field-based meshing.}
% \cite{vaxman16} provides a review of methods for computing direction fields on surfaces.
% The majority of state-of-the-art works on field-based meshing
% compute smooth \emph{cross fields}, i.e., assigning to every point
% of the surface two orthogonal directions indicating the orientation of the mesh.

\subsection{Overview}

In~\autoref{sec:frame_fields} we review the mathematical formalism
behind parametrizations, their Jacobians and frame fields.
We define the integrability property for frame fields
and show how they are equivalent to seamless parametrizations.
In~\autoref{sec:algebraic_representation},
we review the algebraic representation for our frames,
which is the two-dimensional version of the odeco tensors
proposed by~\cite{palmer20a}.
In~\autoref{sec:meshability}, we present two contributions:
(a) an expression for the sensitivity of a higher-order tensor's
eigenvectors with respect to small perturbations of the tensor,
and we use this result to show (b)
an expression of a frame field's Lie bracket
only in terms of the tensor field coefficients
and their derivatives.
In~\autoref{sec:opti} we present a last contribution,
\update{an energy formulation that optimizes for smooth integrable frame fields,
for both the isotropic and anisotropic cases}.
In~\autoref{sec:param} we briefly present
our pipeline to compute a seamless parametrization
from an integrable frame field,
and in~\autoref{sec:results} we demonstrate the effectiveness of the algorithm.

\section{Integrable frame fields} \label{sec:frame_fields}

\paragraph{Seamless parametrizations.}
The idea of parametrization-based quadrilateral meshing
is to map a coordinate system onto the domain to be meshed.
The coordinate lines of this map then provide, at least locally,
a quadrilateral mesh on the domain.
Consider a planar domain $\Omega$; we wish to parametrize $\Omega$
by assigning to each point $\vb{p}$ of $\Omega$
a pair of coordinates $\qty(u(\vb{p}), v(\vb{p}))$.
In general, this parametrization cannot be defined as a global continuous function;
\emph{cuts} need to be introduced to obtain a disk topology
on which $u$ and $v$ can be defined continuously.
The parametrization is said to be \emph{seamless}
if across every cut, the coordinates transform through a rigid rotation
of some multiple of $\SI{90}{\degree}$;
if $u'$ and $v'$ are the coordinates on the other side of the cut,
then the transformation is seamless if
\begin{equation} \label{eq:rigid_rotation}
    \begin{pmatrix} u' \\ v' \end{pmatrix} =
    {\underbrace{\begin{pmatrix} 0 & 1 \\ -1 & 0\end{pmatrix}}_{\vb{R}}}^k
    \begin{pmatrix} u \\ v \end{pmatrix} + 
    \begin{pmatrix} s \\ t \end{pmatrix},
\end{equation}
for some $k \in \{0,1,2,3\}$ and fixed translation $(s,t)$.
If, on top of being seamless,
the translation $(s,t)$ is integer and the singularities lie at integer coordinates,
then the parametrization is called an \emph{integer-grid map} and exactly corresponds to a quadrilateral mesh.

Computing seamless parametrizations is a challenging task 
due to the discrete nature of the cut graph involved,
which is unknown a priori.
Instead, most existing works focus on computing the Jacobian $(\grad{u}, \grad{v})$ of a parametrization.
Let $(\vb{d}u, \vb{d}v)$ be a pair of vector fields on $\Omega$;
they form the Jacobian of a parametrization
(and are said to be \emph{integrable}) if they are curl-free:
\begin{equation} \label{eq:curlfree}
    \curl{\vb{d}u} = \vb{0}, \quad \curl{\vb{d}v} = \vb{0},
\end{equation}
and they transform through a \SI{90}{\degree}-multiple rotation across cuts
(which amounts to differentiating~\eqref{eq:rigid_rotation}):
\begin{equation} \label{eq:rigid_rotation_jacobian}
    \begin{pmatrix} \vb{d}u' \\ \vb{d}v' \end{pmatrix} =
    \vb{R}^k \begin{pmatrix} \vb{d}u \\ \vb{d}v \end{pmatrix}.
\end{equation}
If these properties are verified then the vector fields $(\vb{d}u, \vb{d}v)$
are essentially equivalent to a seamless parametrization,
up to a global rigid rotation of the coordinate map.
Computing a parametrization's Jacobian is a more doable task,
provided one can appropriately encode the symmetries in~\eqref{eq:rigid_rotation_jacobian};
we elaborate on that in~\autoref{sec:algebraic_representation}.

\paragraph{Frame fields.}
Consider now the inverse of the parametrization $\vb{r}$
that maps a pair of coordinates to point of the domain: $\vb{p} = \vb{r}(u,v)$.
This map can only be defined locally since the parametrization can
map different points to the same coordinates.
The Jacobian matrix of $\vb{r}$ defines the \emph{coordinate vectors} $\vb{u}$ and $\vb{v}$:
\begin{equation}
    \vb{J}_{\vb{r}} = 
    \begin{pmatrix} 
        \displaystyle \pdv{\vb{r}}{u} & \displaystyle \pdv{\vb{r}}{v} 
    \end{pmatrix}
    = \begin{pmatrix} \vb{u} & \vb{v} \end{pmatrix},
\end{equation}
forming a \emph{coordinate frame} $\vb{F} = \begin{pmatrix} \vb{u} & \vb{v} \end{pmatrix}$.
Since $\vb{r}$ is the inverse of the parametrization,
their Jacobians are inverse of one another, meaning that the coordinate frame
is the inverse of the parametrization's Jacobian:
\begin{equation}
    \vb{J}_{\vb{r}^{-1}} = \qty(\vb{J}_{\vb{r}})^{-1} \quad \Rightarrow \quad
    \begin{pmatrix} \grad{u} \\ \grad{v} \end{pmatrix} =
    \vb{F}^{-1}.
\end{equation}
Note that, looking at the off-diagonal terms in $\vb{F}^{-1} \vb{F} = \vb{I}$,
we have $\grad{u}\cdot\vb{v} = \grad{v}\cdot\vb{u} = 0$,
meaning that the coordinate vectors $\vb{u},\vb{v}$
are orthogonal to the gradients of $v,u$, respectively,
and therefore parallel to the isolines of $v,u$ respectively.
This shows how the frame field corresponds to a quad mesh
through discrete isolines of the coordinate map $u,v$.
Looking at the diagonal terms, we have $\grad{u}\cdot\vb{u} = \grad{v}\cdot\vb{v} = 1$,
which indicates that the integer isolines are spaced out according to $\norm{\vb{u}}$
and $\norm{\vb{v}}$.
\begin{figure}
    \centering
    \includegraphics[width=.8\linewidth]{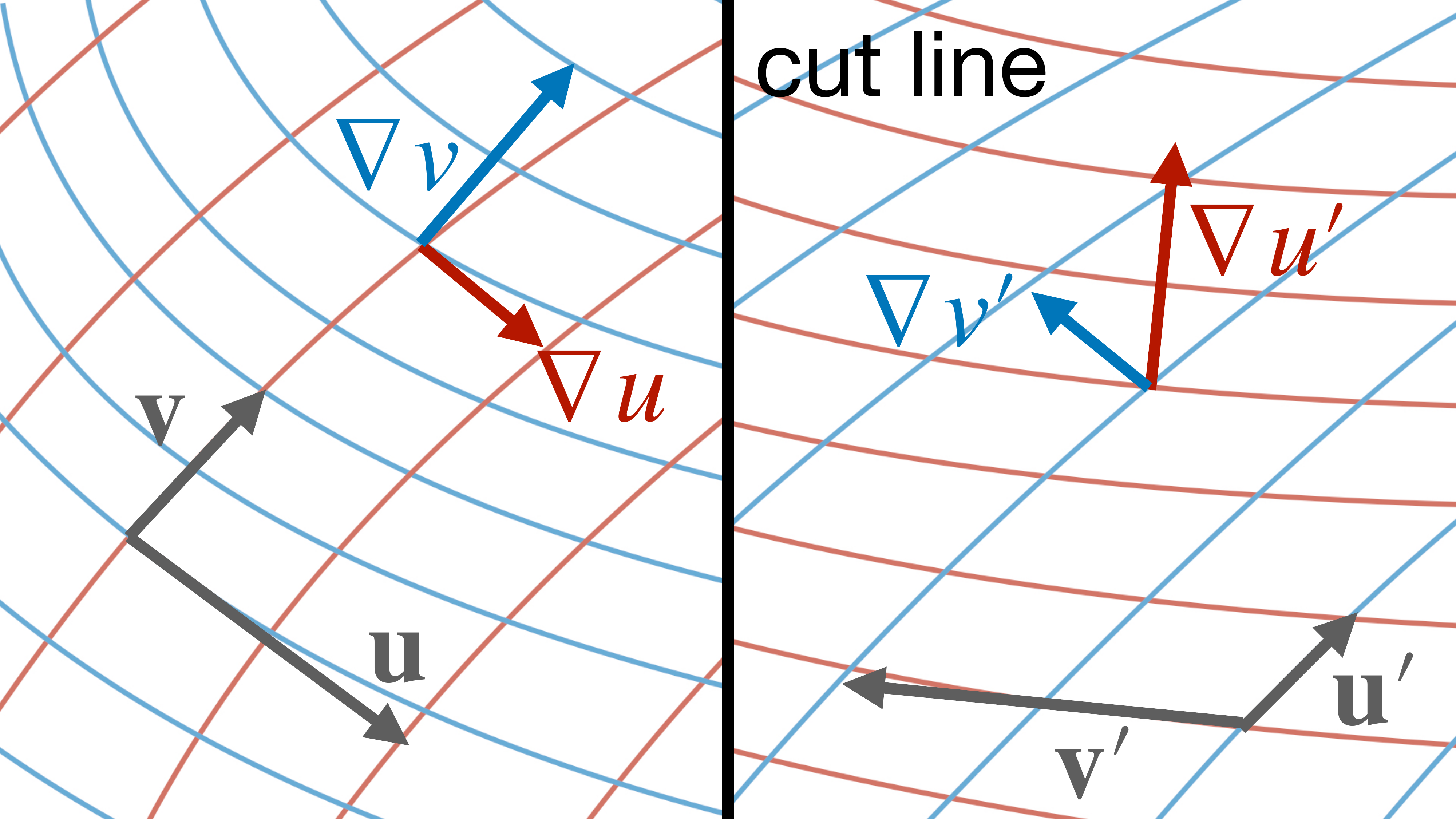}
    \caption{A seamless parametrization across a cut.
    The isolines of the coordinate maps $u(\vb{p})$ and $v(\vb{p})$
    are depicted in red and blue, respectively.
    Notice how the gradients $\grad{u},\grad{v}$ and
    the frame vectors $\vb{u},\vb{v}$ undergo a \SI{90}{\degree} rotation
    as they cross the cut.
    }
    \label{fig:param}
\end{figure}
We illustrate the introduced concepts and notations 
on~\autoref{fig:param}.

\paragraph{Orthogonal case.}
In this work we are interested in frames that are orthogonal: $\vb{u} \perp \vb{v}$.
Let $\vb{\hat{u}} = \vb{u} / \norm{\vb{u}} $ and $\vb{\hat{v}} = \vb{v}/\norm{\vb{v}}$;
the frame matrix $\vb{F}$ is then diagonal in the local orthonormal basis $(\vb{\hat{u}},\vb{\hat{v}})$:
\begin{equation}
    \begin{pmatrix} \grad{u} \\ \grad{v} \end{pmatrix} =
    \begin{pmatrix} \norm{\vb{u}} & 0 \\ 0 & \norm{\vb{v}} \end{pmatrix}^{-1} =
    \begin{pmatrix} 1/\norm{\vb{u}} & 0 \\ 0 & 1/\norm{\vb{v}} \end{pmatrix}.
\end{equation}
The Jacobian of the coordinate map can thus be computed by simply scaling the frame's coordinate vectors:
$\grad{u} = \vb{u}/\norm{\vb{u}}^2$ and $\grad{v} = \vb{v}/\norm{\vb{v}}^2$.

\paragraph{Frame field integrability.}
Several previous works, e.g.,~\cite{diamanti15}, 
build upon the curl-free condition~\eqref{eq:curlfree} and compute
parametrizations by computing vector fields $(\vb{d}u, \vb{d}v)$.
However, in our case we want to be able to express the integrability condition
directly in terms of our implicit frame representation,
and expressing the individual curls of $\vb{d}u$ and $\vb{d}v$
is not possible as the two vectors are mixed in a single algebraic representation.
% which is not possible with the individual curls of $\vb{du}$ and $\vb{dv}$.
Instead, we compute a frame field $\vb{F}$ and rely on the analogous statement:
a frame field $\vb{F} = \begin{pmatrix} \vb{u} & \vb{v} \end{pmatrix}$
guides a parametrization (and is \emph{integrable})
if its $\emph{Lie bracket}$ $\qty[\vb{u},\vb{v}]$ vanishes:
\begin{equation} \label{eq:meshability}
    \qty[\vb{u},\vb{v}] = \grad_{\vb{u}}{\vb{v}} - \grad_{\vb{v}}{\vb{u}} = \vb{0},
\end{equation}
where $\grad_{\vb{u}}$ and $\grad_{\vb{u}}$ are directional derivatives,
and this parametrization is seamless if the frame vectors transform
through $\SI{90}{\degree}$ rotations across cuts,
as in~\eqref{eq:rigid_rotation_jacobian}.
We will see in~\autoref{sec:meshability} how the Lie bracket
can be expressed in terms of the implicit frame representation.

\section{Algebraic representation of frames} \label{sec:algebraic_representation}
Due to the topology of the domain $\Omega$
and the presence of singularities in the frame field,
it is not possible to define a globally continuous frame field $(\vb{u}, \vb{v})$.
One needs to introduce \emph{cuts} in the domain
across which the frame vectors are symmetric according
the rotations defined by~\eqref{eq:rigid_rotation_jacobian}.
\begin{figure}
    \centering
    \includegraphics[trim={0 4cm 19cm 4cm},clip,width=.75\linewidth]{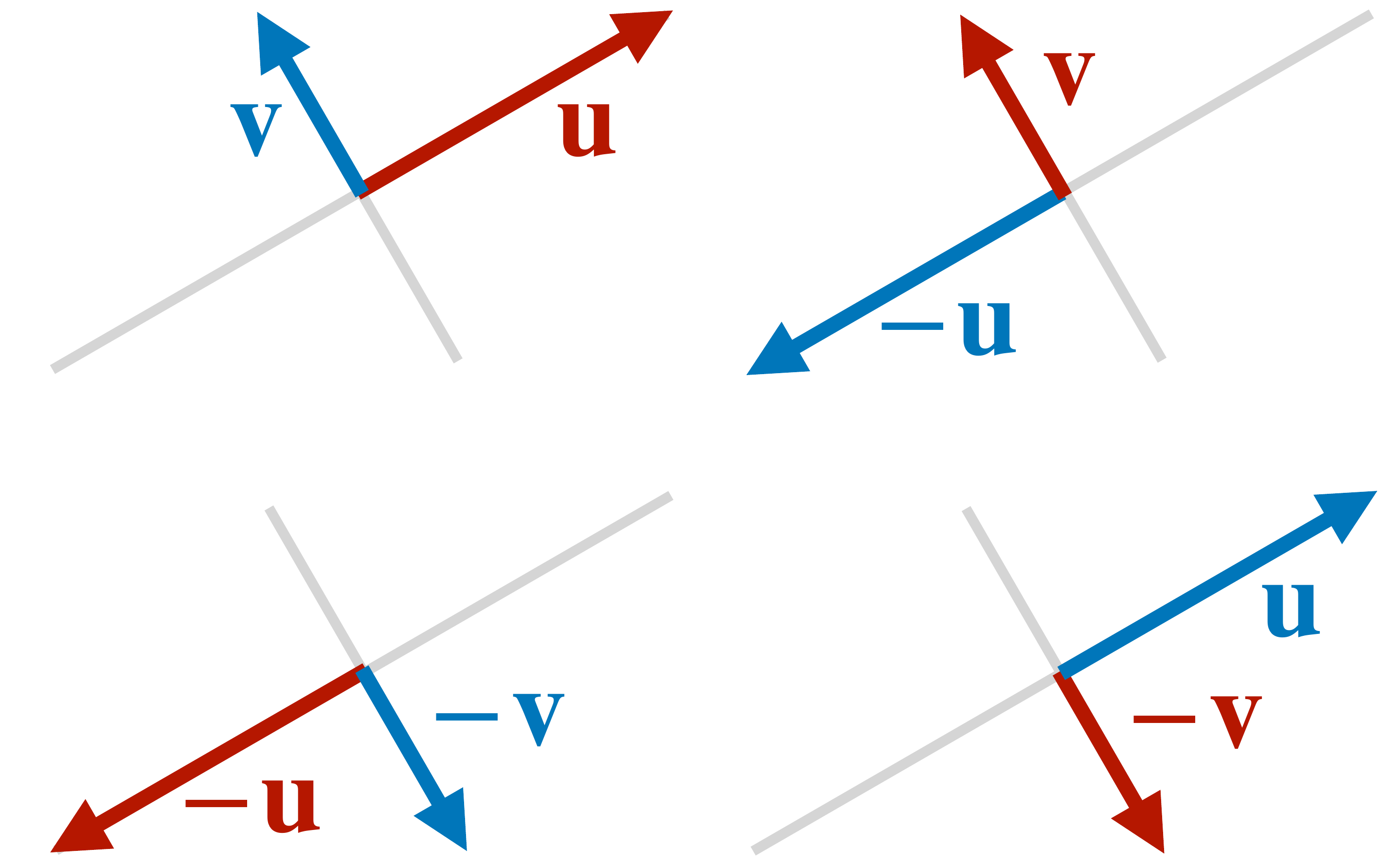}
    \caption{Equivalence class of the \SI{90}{\degree} rotations
    of an orthogonal frame $(\vb{u},\vb{v})$.}
    \label{fig:frames}
\end{figure}
In order to compute a frame field without the need to introduce cuts,
we extend the notion of frame such that it is invariant up to \SI{90}{\degree} rotations:
\begin{equation} \label{eq:permutations}
    \vb{F} = \qty{ (\vb{u},\vb{v}), (\vb{v},-\vb{u}), (-\vb{u},-\vb{v}), (-\vb{v},\vb{u}) }.
\end{equation}
These symmetries are illustrated in~\autoref{fig:frames}.
Given this equivalence class, one needs to introduce an algebraic representation
that unambiguously represents a frame and supports arithmetic operations.
To this end we resort to the class of two-dimensional orthogonally decomposable (or Odeco) tensors,
which were introduced by Palmer et al.~\cite{palmer20a} in the three-dimensional case.
In this section we briefly lay out the theory of odeco tensors,
and refer the reader to the paper for more details.

\paragraph{Odeco tensors.}
\update{Although the odeco theory applies to tensors of arbitrary order,
we restrict the definitions to fourth-order tensors,
as they are sufficient to represent frames.
Let $S^4(\bb{R}^n)$ be the space of $n\times n \times n \times n$ \emph{fully symmetric tensors},
i.e., their real entries $T_{i_1,i_2,i_3,i_4}$ are invariant up to permutations of the indices $i_1,i_2,i_3,i_4$.
$n$ is the \emph{dimension} of the tensor and corresponds to the dimension of the frame we wish to represent (2 or 3).
A combinatorial inspection shows that tensor $\vb{T}$ has 5 independent components for $n=2$,
and 15 components for $n=3$.}

Robeva et al.~\cite{robeva16} have studied a special class of tensors $\vb{T}$
that are said to be \emph{orthogonally decomposable}, or \emph{odeco} for short,
if they can be written as
\begin{equation} \label{eq:odeco}
    \vb{T} = \lambda_1 \vb{v}_1^{\otimes 4} + \dots + \lambda_n \vb{v}_n^{\otimes 4},
\end{equation}
where $\vb{v}_1, \dots, \vb{v}_n$ form an orthonormal basis of $\bb{R}^n$.
Note that for second-order tensors, $S^2(\bb{R}^n)$ is the space of real symmetric matrices;
these are always orthogonally decomposable according to the spectral theorem.
Higher-order tensors ($\geq 3$) however,
admit rank-one decompositions with more than $n$ terms.

The notion of eigenvector can be generalized to higher-order tensors:
$\vb{w}$ is an eigenvector of $\vb{T}$ with eigenvalue $\lambda$ if
\begin{align}
    &\vb{T}\vb{w}^3 = \lambda \vb{w}, \quad \text{or, in Einstein notation,} \label{eq:eigenvector} \\
    &T_{i,j_2,j_3,j_4} w_{j_2} w_{j_3} w_{j_4} = \lambda w_i. \nonumber
\end{align}
One can easily see that, for an odeco tensor $\vb{T}$,
the vectors $\vb{v}_k$ in~\eqref{eq:odeco} are eigenvectors of $\vb{T}$ with corresponding eigenvalue $\lambda_k$.

As odeco tensors only form a small subset of the space of fully symmetric tensors,
one needs to characterize what makes a tensor odeco.
The major result of Robeva et al.~\cite{robeva16} is the fact
that odeco tensors form an algebraic variety defined
by a set of homogeneous quadratic equations in the tensor coefficients.
More precisely, $\vb{T}$ is odeco if the contraction $(\vb{T} * \vb{T})$ defined by
\begin{equation}
    (\vb{T} * \vb{T})_{i_1,i_2,i_3,j_1,j_2,j_3} = T_{i_1,i_2,i_3,s} T_{j_1,j_2,j_3,s}
\end{equation}
is a fully symmetric tensor, i.e.,
\begin{equation} \label{eq:variety}
    \vb{T}*\vb{T} \in S^{6}(\bb{R}^n).
\end{equation}

\paragraph{Frames as odeco tensors.}
\update{To represent a 2D orthogonal frame in a way that is invariant to the permutations in~\eqref{eq:permutations},
we use an odeco tensor of which the eigenvalues and eigenvectors, as defined in~\eqref{eq:eigenvector},
match the frame directions and sizes.}
Specifically, let $\vb{u},\vb{v}$ be the frame vectors,
$\lambda,\mu$ their norms and $\vb{\hat{u}}, \vb{\hat{v}}$ their normalization such that
$\vb{u} = \lambda\vb{\hat{u}},\ \vb{v} = \mu\vb{\hat{v}}$.
Then the corresponding tensor is defined as
\begin{equation} \label{eq:odeco_tensor}
    \vb{T} = \lambda \vb{\hat{u}}^{\otimes 4} + \mu \vb{\hat{v}}^{\otimes 4}.
\end{equation}
Notice how this tensor remains the same when plugging in
any of the rotations of~\eqref{eq:permutations}. % \notealex{3.6 does not define permutations}
We can now justify the choice for a fourth-order tensor:
an even order is required for the tensor to be invariant to the signs of the vectors,
and order $2$ cannot be chosen since, in the case $\lambda=\mu$,
the tensor degenerates to a multiple of the identity matrix, $\vb{T} = \lambda\vb{I}$,
and then any vector of $\bb{R}^2$ is an eigenvector of $\vb{T}$
and one loses the direction vectors $\vb{\hat{u}}$, $\vb{\hat{v}}$.
Order $4$ is therefore the lowest possible order for our purpose.

\begin{figure}
    \centering
    \includegraphics[width=.45\linewidth]{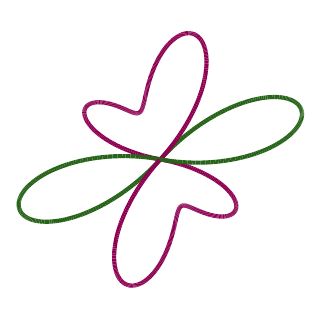}
    \includegraphics[width=.45\linewidth]{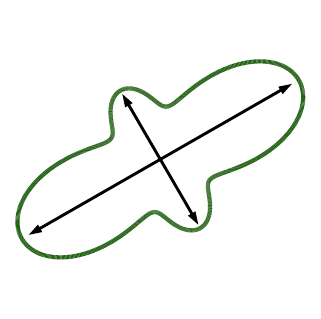}
    \vspace{-1em}
    \caption{\small Polynomial representation $p_{\vb{T}}(\theta)$ of (left) an arbitrary tensor,
    and (right) an odeco tensor, with the corresponding frame vectors.
    Green represents positive values for $p_{\vb{T}}(\theta)$
    and magenta represents negative values.}
    \label{fig:poly_examples}
\end{figure}

\paragraph{Polynomial representation of tensors.}
There is a one-to-one correspondence between 4th-order fully symmetric tensors
of dimension $n$, and degree 4 homogeneous polynomials of $n$ variables, given by
\begin{equation} \label{eq:pT}
    p_{\vb{T}}(x_1,\dots,x_n) = T_{i_1,i_2,i_3,i_4} x_{i_1} x_{i_2} x_{i_3} x_{i_4}.
\end{equation}
Since it is homogeneous, one can restrict this polynomial to the sphere $\bb{S}_{n-1}$, since
\begin{equation}
    p_{\vb{T}}(\vb{x}) = \norm{\vb{x}}^4 \ p_{\vb{T}}\qty(\frac{\vb{x}}{\norm{\vb{x}}}).
\end{equation}
We can therefore define a function $p_{\vb{T}}(\theta)$
such that $p_{\vb{T}}(\vb{x}) = \norm{\vb{x}}^4 \ p_{\vb{T}}(\theta)$;
this function is periodic with period $\pi$, since $p_{\vb{T}}(\vb{x}) = p_{\vb{T}}(-\vb{x})$.
Such a function can be conveniently visualized by taking a unit circle and virtually stretching it
according to the value that $p_{\vb{T}}$ takes on the circle;
formally, one draws the parametric curve given by $r(\theta) = p_{\vb{T}}(\theta)$
for $\theta \in [0,2\pi[$.
We show this on~\autoref{fig:poly_examples}.

\begin{figure}
    % \centering
    \hspace{-.03\linewidth}
    \begin{subfigure}{.19\linewidth}
        \includegraphics[width=\linewidth]{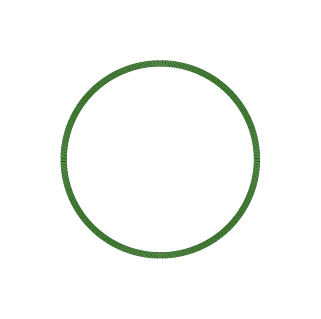}
        \caption*{$\frac{1}{\sqrt{2\pi}}$}
    \end{subfigure}
    \begin{subfigure}{.19\linewidth}
        \includegraphics[width=\linewidth]{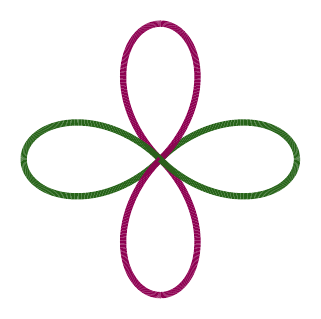}
        \caption*{$\frac{1}{\sqrt{\pi}}\cos(2\theta)$}
    \end{subfigure}
    \begin{subfigure}{.19\linewidth}
        \includegraphics[width=\linewidth]{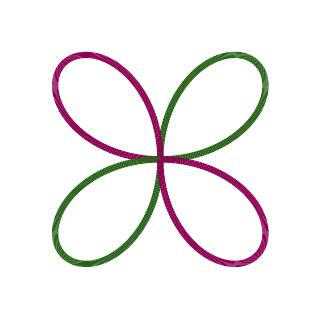}
        \caption*{$\frac{1}{\sqrt{\pi}}\sin(2\theta)$}
    \end{subfigure}
    \begin{subfigure}{.19\linewidth}
        \includegraphics[width=\linewidth]{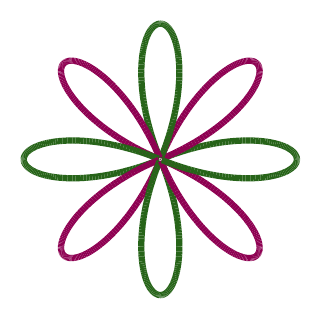}
        \caption*{$\frac{1}{\sqrt{\pi}}\cos(4\theta)$}
    \end{subfigure}
    \begin{subfigure}{.19\linewidth}
        \includegraphics[width=\linewidth]{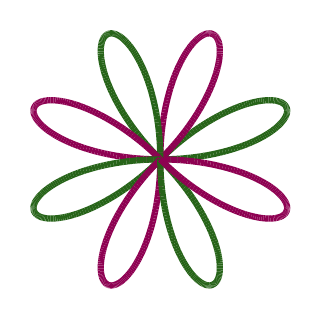}
        \caption*{$\frac{1}{\sqrt{\pi}}\sin(4\theta)$}
    \end{subfigure}
    \caption{\small Orthonormal basis of circular harmonics
    used to represent odeco tensor polynomials.
    Green represents positive values for $p_{\vb{T}}(\theta)$
    and magenta represents negative values.}
    \label{fig:sph_basis}
\end{figure}

A natural way to encode this polynomial is to decompose it into an orthonormal
basis of functions on the sphere $\bb{S}_{n-1}$.
Considering the two-dimensional case $n=2$,
$p_{\vb{T}}(\theta)$ can be decomposed into the Fourier series
% If we now restrict to the 2D case $n=2$,
% the five basis functions are sometimes called \emph{circular harmonics}
% (in reference to the well-known spherical harmonics in 3D),
% and correspond to a Fourier series.
% Let $p_{\vb{T}}(\theta)$ denote the restriction of $p_{\vb{T}}$ to the unit circle;
% this function can be decomposed as
\begin{equation} \label{eq:pT_theta}
    \begin{aligned}
        p_{\vb{T}}(\theta) = \frac{1}{\sqrt{2\pi}} \ q_0 &+ \frac{1}{\sqrt{\pi}} \cos(2\theta) \ q_1 + \frac{1}{\sqrt{\pi}} \sin(2\theta) \ q_2 \\
         &+ \frac{1}{\sqrt{\pi}} \cos(4\theta) \ q_3 + \frac{1}{\sqrt{\pi}} \sin(4\theta) \ q_4,
    \end{aligned}
\end{equation}
where $\vb{q} = (q_0, \dots, q_4)$ fully describes the polynomial and its corresponding tensor.
These basis functions, shown in~\autoref{fig:sph_basis},
correspond to two-dimensional spherical harmonics
and are sometimes called \emph{circular harmonics}.
%\notealex{Aren't they spherical harmonics of dimension 2?}
The number of five is not surprising since the original tensor has five independent components:
$T_{1111}$, $T_{1112}$, $T_{1122}$, $T_{1222}$ and $T_{2222}$.
The change of basis from $\vb{T}$ to $\vb{q}$ is done through a linear transformation,
\update{which can be derived by matching expressions~\eqref{eq:pT} and~\eqref{eq:pT_theta}}.

Eigenvectors and eigenvalues of higher-order tensors, defined in~\eqref{eq:eigenvector},
can be interpreted in the polynomial representation through the following property:
$\vb{w}$ is an eigenvector of $\vb{T}$ with eigenvalue $\lambda$ if and only if
\begin{equation}
    \grad{p_{\vb{T}}}(\vb{w}) = 4 \ \lambda \ \vb{w}.
\end{equation}
In other words, the eigenvectors of $\vb{T}$ correspond to the stationary points of $p_{\vb{T}}(\theta)$.

\paragraph{Algebraic variety in the 2D case.}
A 2D fourth-order tensor is completely defined by a set of five coefficients $(q_0,\dots,q_4)$.
This tensor corresponds to a frame if the tensor is odeco,
i.e., it lies on the algebraic variety defined by the set of quadratic equations~\eqref{eq:variety}.
In the 2D case \update{($n=2$),
we can write out this set of equations and change the basis to $\vb{q}$,
and derive the following set of linearly independent equations that define the variety}:
\begin{equation} \label{eq:odeco_alg}
    \left\{ \begin{aligned}
        c_1(\vb{q}) &= & q_0^2 - 18(q_3^2 + q_4^2) &= 0 \\
        c_2(\vb{q}) &= & \sqrt{2}q_0q_1 - 6q_1q_3 - 6q_2q_4 &= 0 \\
        c_3(\vb{q}) &= & \sqrt{2}q_0q_2 - 6q_1q_4 + 6q_2q_3 &= 0
    \end{aligned} \right.
\end{equation}

\paragraph{Area of a frame.}
The optimization we perform on the frame field
requires to normalize the integrability condition
to make it independent of the size of the frame.
We choose to do this normalization with the \emph{area}
of the frame, i.e., the product of the sizes $a = \lambda\mu$.
This can be interpreted as the local area of the quad elements
in the frame's neighborhood.
One can show that it can be expressed only in terms of the circular harmonics
coefficients $\vb{q}$:
% To preserve the one-to-one correspondence between frames and odeco tensors,
% the sizes $\lambda$ and $\mu$ in~\eqref{eq:odeco_tensor} must remain positive.
% To impose this during the optimization we will introduce a barrier term \notealex{I don't think this is the best way to present a. It's not really used like a barrier term, more like a normalization of the energy we want to minimize. And this normalization happens to act as a barrier. I would present it as a normalization term, and then state that it happens to act as a barrier if we start from positive initialization --as you are doing in section 5.--}
% involving the product $a = \lambda\mu$. This can be interpreted
% as the local area of the quad elements in the frame's neighborhood.
% One can show that it can be expressed only in terms of the circular harmonics coefficients $\vb{q}$:
\begin{equation}
    a(\vb{q}) = \lambda\mu = \frac{1}{\pi} \qty( \frac{8}{9} q_0^2 - (q_1^2 + q_2^2) ).
\end{equation}
We show later in Section~\ref{sec:opti} how this expression is used to normalize the integrability condition.

\paragraph{Isotropic case.}
A frame is \emph{isotropic},
i.e., its vectors have equal length $\lambda=\mu$,
if its degree 2 coefficients are zero: $q_1 = q_2 = 0$.
The algebraic variety reduces to a single quadratic equation
$c_1(\vb{q}) = 0$.
When the degree 2 coefficients are zero,
the tensor eigenvectors are always orthogonal,
even if the tensor is not odeco.
\begin{figure}
    \centering
    \includegraphics[width=.195\linewidth]{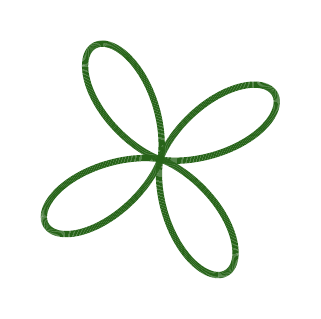}%\hspace{-1em}
    \includegraphics[width=.195\linewidth]{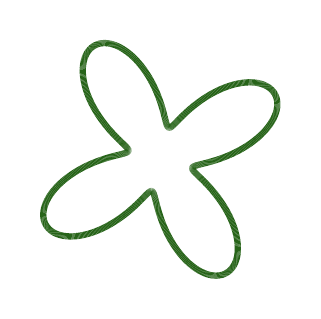}%\hspace{-1em}
    \includegraphics[width=.195\linewidth]{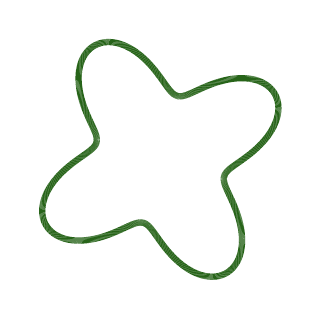}%\hspace{-1em}
    \includegraphics[width=.195\linewidth]{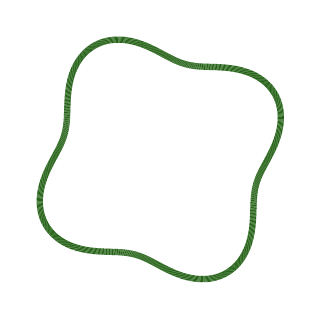}%\hspace{-1em}
    \includegraphics[width=.195\linewidth]{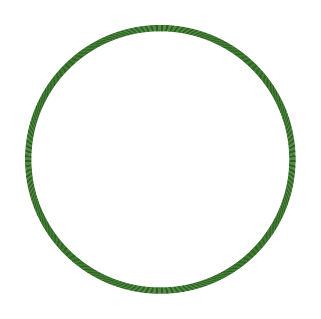}
    \caption{\small Range of isotropic tensors, the middle one being odeco.
    All tensors have a set of 4 eigenvectors with the same eigenvalues.}
    \label{fig:sph_iso}
\end{figure}
A range of isotropic tensors is illustrated on~\autoref{fig:sph_iso}.
We see that in the general non-odeco case, this class has an additional degree of freedom
depending on how close it is to a sphere.
We explain in~\autoref{sec:results}
how this extra degree of freedom is beneficial as it allows
to represent singularities.

\section{Integrability condition} \label{sec:meshability}
Since there is a one-to-one correspondence between frames (as defined in~\eqref{eq:permutations})
and orthogonally decomposable (odeco) tensors with positive eigenvalues (as defined in~\eqref{eq:odeco_tensor}),
one can express the integrability condition~\eqref{eq:meshability}
solely in terms of the odeco tensor coefficients and their spatial derivatives.
In this section we show how we derive the closed-form expression for the Lie bracket.

Consider a frame $\vb{F} = \qty{\vb{R}^k (\vb{u},\vb{v}), \ k=0,1,2,3}$,
\update{where $\vb{R}$ performs a \SI{90}{\degree} rotation as defined in~\eqref{eq:rigid_rotation}
and illustrated in~\autoref{fig:frames}.}
Every pair of vectors in $\vb{F}$ has the same Lie bracket,
which we denote $\rm{Lie}(\vb{F})$:
\begin{equation}
    \rm{Lie}(\vb{F}) = [\vb{R}^k\vb{u},\vb{R}^k\vb{v}], \quad k=0,1,2,3.
\end{equation}
First we write the Lie bracket $\qty[\vb{u},\vb{v}]$ in index notation:
\begin{equation}
    \qty[\vb{u},\vb{v}]_i = u_\alpha \pdv{v_i}{x_\alpha} - v_\alpha \pdv{u_i}{x_\alpha}.
\end{equation}
\update{Let $\vb{T}$ denote the odeco tensor corresponding to $\vb{F}$.}
We apply the chain rule to express the spatial derivatives on the tensor coefficients:
\begin{equation} \label{eq:lie_chain_rule}
    \qty[\vb{u},\vb{v}]_i = u_\alpha \pdv{v_i}{T_{j_1j_2j_3j_4}} \pdv{T_{j_1j_2j_3j_4}}{x_\alpha} - v_\alpha \pdv{u_i}{T_{j_1j_2j_3j_4}} \pdv{T_{j_1j_2j_3j_4}}{x_\alpha}.
\end{equation}
In the remaining of the section we show how
(a) the sensitivity terms $\pdv*{v_i}{T_j}$ and $\pdv*{u_i}{T_j}$ are derived,
and (b) how the Lie bracket is completely expressed in terms of the tensor coefficients $\vb{T}$.

\paragraph{Eigenvalue sensitivity for tensors.}
% Let $\vb{T}$ be the odeco tensor formed by $\vb{u}$ and $\vb{v}$,
% and $\vb{q}$ its coefficients in the basis of circular harmonics.
Recall that $\vb{u},\vb{v}$ are eigenvectors of $\vb{T}$ as defined by~\eqref{eq:eigenvector}.
Finding the sensitivity terms $\pdv*{v_i}{T_j}$ and $\pdv*{u_i}{T_j}$
can be done by analyzing how the eigenvectors $\vb{u},\vb{v}$
change when adding an infinitesimal perturbation $\delta\vb{T}$ to the tensor.
This analysis is well-known for matrices as the \emph{eigenvalue perturbation problem},
but can be generalized to higher-order tensors, provided they are odeco.
We leave the complete derivation in appendix~\ref{app:sensitivity}
and retain the main result:
if $(\lambda,\vb{\hat{w}})$ is an eigenpair of odeco tensor $\vb{T}$,
then the sensitivity of $\vb{w} = \lambda\vb{\hat{w}}$
with respect to a perturbation $\delta\vb{T}$ is given by
\begin{equation}
    \delta\vb{w} = (\delta\vb{T}) \vb{\hat{w}}^3.
\end{equation}
% \notealex{There is something wrong here i think. This is working only if w = lambda*ŵ with lambda eigenvalue.
% When saying w is an eigenvector, it is not excluding the case were $||w|| = N,$ with N!= lambda.
% Either work with ŵ unitary eigenvector and lambda associated eigenvalue, either with w eigenvector such as $||w|| = lambda$, with lambda associated eigenvalue.}
From this result we find that the partial derivatives are given by
\begin{equation} \label{eq:sensitivity}
    \pdv{w_i}{T_{j_1j_2j_3j_4}} = \delta_{ij_1} \hat{w}_{j_2} \hat{w}_{j_3} \hat{w}_{j_4},
\end{equation}
where $\delta_{ij}$ is the Kronecker delta symbol.
Note that an eigenvector's sensitivity only depends on itself
and not on the other eigenvectors of the odeco tensor.

\paragraph{Lie bracket in terms of tensor representation.}
We now wish to find an expression for the Lie bracket~\eqref{eq:lie_chain_rule}
so that it only depends on the tensor coefficients $\vb{T}$
and not on the frame directions $\vb{u},\vb{v}$.
Plugging in the sensitivity result~\eqref{eq:sensitivity} yields
\begin{equation}
    \qty[\vb{u},\vb{v}]_i = \qty( \norm{\vb{u}} \hat{u}_\alpha \hat{v}_{j_2}\hat{v}_{j_3}\hat{v}_{j_4} 
    - \norm{\vb{v}} \hat{v}_k \hat{u}_{j_2}\hat{u}_{j_3}\hat{u}_{j_4}) \pdv{T_{ij_2j_3j_4}}{x_\alpha}.
\end{equation}
We see that an expression close to the odeco tensor definition~\eqref{eq:odeco_tensor} starts to appear.
The last ingredient is to note that $\vb{\hat{u}}$ and $\vb{\hat{v}}$ are related by a $\SI{90}{\degree}$ rotation,
which can be written an $\hat{v}_i = \epsilon_{ij}\hat{u}_j$
and $\hat{u}_i = -\epsilon_{ij}\hat{v}_j$, where $\epsilon_{ij}$ is the two-dimensional Levi-Civita symbol.
Plugging this into the Lie bracket expression and identifying the tensor coefficients yields
\begin{equation}
    \rm{Lie}(\vb{T})_i = \epsilon_{j_2k_2} \epsilon_{j_3k_3} \epsilon_{j_4k_4}
    T_{\alpha k_2 k_3 k_4} \pdv{T_{ij_2j_3j_4}}{x_\alpha}.
\end{equation}
As the tensor coefficients $\vb{T}$ are a linear transformation
of the polynomial coefficients $\vb{q}$,
one can express the Lie bracket in terms of $\vb{q}$ only:
\begin{equation} \label{eq:lie_tensor}
    \rm{Lie} (\vb{q})_i = C_{ijk\alpha} \ q_k \ \pdv{q_j}{x_\alpha},
\end{equation}
where the $C_{ijk\alpha}$ are known coefficients.
\update{This expression is a key result of this work
as it allows to compute an integrability metric
that does not involve the frame vectors,
but instead a quadratic function of the implicit tensor representation.}

\section{Optimization} \label{sec:opti}
We design an optimization formulation that aims at
making the Lie bracket~\eqref{eq:lie_tensor} as close to zero as possible,
with odeco tensors~\eqref{eq:odeco_alg} having positive sizes ($\lambda,\mu > 0$).
To do so we define the following energy functionals on the domain $\Omega$:
\begin{align}
    E_\rm{Lie}[\vb{q}] &= \int_\Omega \frac{\norm{\rm{Lie}(\vb{q})}^2}{a(\vb{q})^2} \ \dd{\vb{x}}, \\
    E_\rm{odeco}[\vb{q}] &= \sum_{i=1}^3 \int_\Omega c_i(\vb{q})^2 \ \dd\vb{x}, \\
    E_\rm{D}[\vb{q}] &= \sum_{j=1}^5 \frac{1}{2} \int_\Omega \norm{\grad{q_j}}^2 \ \dd{\vb{x}}.
\end{align}
The normalized Lie bracket energy $E_\rm{Lie}$ aims at making the frame field integrable,
the odeco penalty $E_\rm{odeco}$ at keeping the tensors odeco,
and the Dirichlet energy $E_\rm{D}$ strives for smoothness of the frame field
\update{(Palmer et al.~\cite{palmer20a} have shown that, since the $L^2$ distance on $\vb{q}$
corresponds to the $L^2$ distance between the polynomials $p_{\vb{T}}$,
the Dirichlet energy is an adequate proxy for the smoothness of the frame field)}.
The area expression $a(\vb{q})$ in the denominator of the Lie bracket energy
acts as a normalization:
$E_\rm{Lie}$ does not depend on the global scale of the frame field, i.e.,
for a constant factor $\alpha$, $E_\rm{Lie}(\alpha\vb{q}) = E_\rm{Lie}(\vb{q})$.
\update{It also acts as a barrier, preventing the frame sizes $\lambda$ and $\mu$ from becoming zero or negative,
provided the initial solution has positive sizes}.
The Lie bracket energy and the odeco penalty are assembled in a Ginzburg-Landau-like functional,
and the Dirichlet energy acts as a regularizer:
\begin{equation} \label{eq:energy}
    \min_{\vb{q}} E_\kappa[\vb{q}] = 
    (1-\kappa) \ E_\rm{Lie}[\vb{q}] + \kappa \ E_\rm{D}[\vb{q}] + \frac{1}{\epsilon^2} E_\rm{odeco}[\vb{q}].
\end{equation}
Parameter $\epsilon$ has units of length and,
like in the Ginzburg-Landau functional,
controls the size of the singularity neighborhoods
where the frames drift away from the odeco variety.
A schedule is set on $\kappa$ such that it starts at 1,
providing a smooth (but not integrable) frame field,
and is gradually reduced to 0 to enforce integrability.
\update{As there is an infinite space of integrable solutions,
we found that the Dirichlet regularization
prevents the solver from drifting too far away from
smooth solutions.}

\paragraph{Discretization.}
The frame field is discretized on a standard
triangle mesh over $\Omega$.
At each node we store the five circular harmonics
coefficients $\vb{q} = (q_0,\dots,q_4)$.
The coefficients are then linearly interpolated
on the mesh using continuous P1 triangular finite elements.
The energy functionals are then evaluated
using a standard 3-point quadrature rule on the triangles.
% \note{Is 3 points enough? Why?} \notealex{probably not. We can check it's influence on results, but very probably useless to put it higher}
Note that we are using a continuous approximation for the frame field
even though the actual field is not defined at the singularities
and therefore cannot be represented by our discretization.
This justifies our weak enforcement of the odeco constraint
through a penalty term:
it allows the creation of singularities in a way
that does not blow up the integrability energy $E_\rm{Lie}$,
by using tensors which have no preferential directions,
i.e., a ball as in~\autoref{fig:sph_basis}, left.

\begin{figure}
    \centering
    \includegraphics[width=\linewidth]{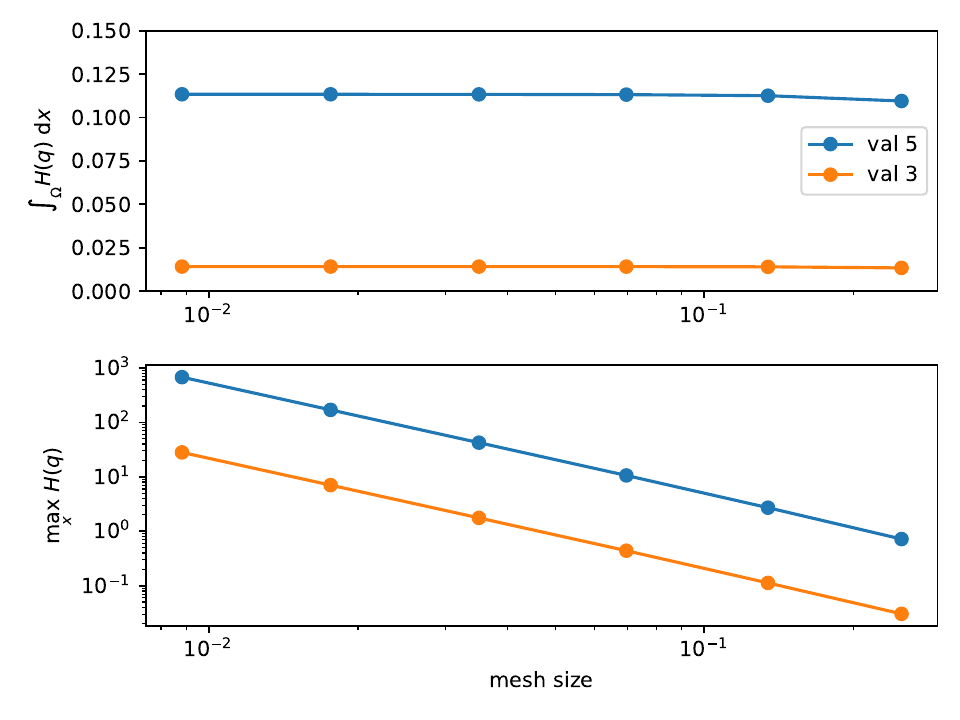}
    \vspace*{-2em}
    \caption{\small
    % \notealex{If there is time would be good to go smaller than $1e-2$ for mesh size ?}
    Convergence of the Lie bracket energy $E_\mathrm{Lie}$ around singularities of valence 5 and 3. $H(\vb{q})$ denotes the integrand of $E_\mathrm{Lie}$: $H(\vb{q}) = \norm{\rm{Lie}(\vb{q})}^2 / a(\vb{q})^2$.}
    \label{fig:convergence}
\end{figure}

\paragraph{Behavior of $E_\mathrm{Lie}$ at singularities.}
Since we are using a continuous representation for the frame field,
a legitimate concern is whether the Lie bracket energy $E_\mathrm{Lie}$
remains bounded around singularities.
To analyze this we compute an exactly integrable frame field
around a singularity of index $i$; this frame field is given in polar coordinates by
\begin{equation}
    \begin{aligned}
        \vb{u}(r,\theta) &= r^i \ \qty( \cos((1-i)\theta) \ \vb{\hat{e}}_r - \sin((1-i)\theta) \ \vb{\hat{e}}_\theta ), \\
        \vb{v}(r,\theta) &= r^i \ \qty( \sin((1-i)\theta) \ \vb{\hat{e}}_r + \cos((1-i)\theta) \ \vb{\hat{e}}_\theta ).
    \end{aligned}
\end{equation}
Notice that the size $r^i$ vanishes at the singularity for $i>0$ (valence 3 and less)
and blows up for $i<0$ (valence 5 and more).
One can check that this frame field indeed has zero Lie bracket.
We represent this frame field on finite element meshes of varying sizes
and evaluate both the total energy $E_\mathrm{Lie}$
as well as the maximum value of the integrand;
this is shown on~\autoref{fig:convergence}.
We see that even though the value of the integrand blows up when refining the mesh,
the energy remains bounded and converges to a fixed value for both singularities.
The energy is higher for a valence 5 singularity than a valence 3 since the size is unbounded.
This property makes it possible for our solver to introduce singularities
whatever the mesh size, if they make the frame field more integrable.

\paragraph{Recovering a frame field.}
The minimization of~\eqref{eq:energy} provides
a field of tensors that is odeco everywhere except
in the vicinity of singularities.
We recover a frame at every node of the mesh in two steps:
first, the tensors are projected onto the odeco
variety using the projector of~\cite{palmer20a},
using a semidefinite relaxation of the exact projection problem.
This operation results in a set of frames
of which we only keep the direction vectors $\vb{\hat{u}},\vb{\hat{v}}$.
In a second step, the sizes of the direction vectors
are computed by contracting the tensor 4 times,
as if they were eigenvectors:
\begin{equation} \label{eq:recover_sizes}
    \lambda = \vb{T}\vb{\hat{u}}^4, \quad \mu = \vb{T}\vb{\hat{v}}^4.
\end{equation}
We found that this approach for recovering sizes
provided better results than using the sizes given by the projector;
this is because the Lie bracket is expressed in terms of the eigenvectors
and eigenvalues of the odeco tensor field.

\update{The optimization procedure to compute an integrable frame field is summarized in Algorithm~\ref{alg:parametrization}.}
\begin{algorithm}
    \caption{\update{Integrable frame field solver}}
    \label{alg:parametrization}
    
    \begin{algorithmic}
        \State \textbf{Input:} A triangle mesh on a planar geometry, with size and/or orientation constraints on boundaries
        \State \textbf{Output:} A smooth integrable frame field\\
        \State On boundaries, assign frames respecting constraints.
        \State In interior, assign zero frames.
        \State Compute $\vb{T}$ and $\vb{q}$ using~\eqref{eq:odeco_tensor}.
        \For{$\kappa$ in $[1, 10^{-1}, 10^{-2}, 10^{-3}, 10^{-4}, 0]$}
            \State $\vb{q} \leftarrow \text{argmin}_{\vb{q}} E_\kappa[\vb{q}]$ using L-BFGS. \eqref{eq:energy}
        \EndFor
        \For{each node}
            \State Recover directions $\vb{\hat{u}}, \vb{\hat{v}}$ using projection of~\cite{palmer20a}.
            \State Recover frame sizes $\lambda,\mu$ using~\eqref{eq:recover_sizes}.
        \EndFor
        \end{algorithmic}
\end{algorithm}

\section{Frame field guided parametrization and meshing} \label{sec:param}
We briefly review how a seamless parametrization
is computed from the frame field $\vb{F}(\vb{x})$ obtained
through the optimization of~\eqref{eq:energy};
for a more detailed treatment we refer to, e.g.,~\cite{bommes09}.

We compute a field-guided seamless parametrization with the following steps:
\begin{description}
    \item[Singularity detection.]
    Singular triangles are identified by calculating
    the turning number of the frames around a given triangle;
    this number is $+1/4$ or $-1/4$ for a singularity of valence 3 and 5, respectively.
    \item[Cut graph.] The triangle mesh is "cut open"
    such that a closed curve cannot turn around a singularity
    without crossing the cut graph. 
    The singular triangles also belong to the cut graph
    (the parametrization is not defined on them).
    \item[Frame vectors assignment.]
    Thanks to the cut graph, two vectors $\vb{u}$ and $\vb{v}$
    can be chosen from each frame to form two continuous vector fields
    $\vb{u}(\vb{x})$ and $\vb{v}(\vb{x})$.
    \item[Integration.]
    The vector fields are integrated to scalar potentials $u(\vb{x})$
    and $v(\vb{x})$ by solving the least-squares problem
    \begin{equation} \label{eq:param_LS}
        \min_{\substack{u(\vb{x}) \\ v(\vb{x})}} 
        \int_\Omega \qty(\norm{\grad{u} - \frac{\vb{u}}{\norm{\vb{u}}^2}}^2
        + \norm{\grad{v} - \frac{\vb{v}}{\norm{\vb{v}}^2}}^2) \ \dd{\vb{x}}.
    \end{equation}
    During this optimization, the parametrization is constrained to be seamless,
    meaning that (i) on the boundary,
    one of the potentials $u$ or $v$ is constant, and
    (ii) along each cut, the matching potential gradients are equal
    (which amounts to impose~\eqref{eq:rigid_rotation_jacobian}).
    These seamlessness constraints can be written as linear constraint
    on the unknowns which are straightforward to impose.
\end{description}

We call the residual of the optimization objective~\eqref{eq:param_LS}
the \emph{integration error}; it quantifies how integrable the frame field was,
and in general how well the parametrization aligns to the prescribed frame field.
Achieving a small integration error means we can adequately control
the size and orientation constraints in the seamless parametrization process.

Different approaches, reviewed earlier in~\autoref{sec:intro},
exist to generate a fully quadrilateral mesh from a seamless parametrization.
In this work we use the quasi-structured quadrilateral mesher of~\cite{reberol21},
which consists in a frontal point insertion coupled with cavity remeshing
and mesh smoothing.

\section{Results and Discussion} \label{sec:results}
To demonstrate the validity of our method,
we compare smooth frame fields against integrable frame fields
for a set of geometries where the frame orientations and sizes
are fixed on the boundaries.
For each frame field we compute a parametrization
as described in the previous section, and measure the integration error.
For the integrable frame field we also generate a quadrilateral mesh
using the previously described frontal mesher.

\paragraph{Parameter settings.}
For each test case we use a fixed odeco parameter $\epsilon$
that has the order of magnitude of the maximum mesh size.
The initial solution of the optimization
is a smooth field of tensors that are not constrained to be odeco, i.e.,
$\kappa=1$ and $\epsilon=\infty$.
Then a schedule is started on the reguarization parameter $\kappa$,
using the following sequence:
$(1, 10^{-1}, 10^{-2}, 10^{-3}, 10^{-4}, 0)$.
The first iteration ($\kappa=1$) corresponds to a pure Ginzburg-Landau functional
(this is the solution we use for smooth frame fields),
and at the last iteration ($\kappa=0$) the regularizer is removed to obtain
the most integrable frame field.
To solve each optimization problem~\eqref{eq:energy}
we use a standard quasi-Newton L-BFGS solver.
The integral calculations are done in parallel over the triangles.
For the following results, the triangular meshes have between 1000 and 4000 nodes.
All results have been obtained in less than 3 minutes
on a \SI{2}{\giga\hertz} Intel Core i5 CPU on 4 threads. 

\begin{figure*}
    \vspace{-1cm}
    \centering
    \hspace*{-.5cm}
    \begin{tabular}{@{}c@{}c|@{}c@{}c@{}c@{}}
        \textbf{Smooth frame field} & \textbf{Smooth param}. & \textbf{Integrable frame field} & \textbf{Integrable param.} & \textbf{Quad mesh}
        \\\hline

\addlinespace

        \multicolumn{5}{l}{(a) Size is 1 at bottom, 2 at top, and varies linearly on the sides.} \\
        % \multicolumn{2}{c}{
        %     \begin{tabular}{@{}c@{}c@{}}
        %         \includegraphics[trim={2cm 2cm 2cm 2cm},clip,width=.2\textwidth]{img/res/square_1_2_smooth_iso_frames.png} %\hspace{-1em}
        %         &\includegraphics[trim={2cm 2cm 2cm 2cm},clip,width=.2\textwidth]{img/res/square_1_2_smooth_iso_param.png} \\ 
        %         \multicolumn{2}{c}{integ. error: 0.0194898}
        %     \end{tabular}
        % }
        % &
        % \multicolumn{3}{|c}{
        %     \begin{tabular}{@{}c@{}c@{}c@{}}
        %         \includegraphics[trim={2cm 2cm 2cm 2cm},clip,width=.2\textwidth]{img/res/square_1_2_iso_frames.png} %\hspace{-1em}
        %         &\includegraphics[trim={2cm 2cm 2cm 2cm},clip,width=.2\textwidth]{img/res/square_1_2_iso_param.png}  %\hspace{-1em}
        %         &\includegraphics[trim={2cm 2cm 2cm 2cm},clip,trim={2cm 2cm 2cm 2cm},clip,width=.2\textwidth]{img/res/square_1_2_qs_integrable.png} \\ 
        %         \multicolumn{2}{c}{(isotropic) integ. error: 0.00313874} & \\ 
        %         \includegraphics[trim={2cm 2cm 2cm 2cm},clip,width=.2\textwidth]{img/res/square_1_2_aniso_frames.png} %\hspace{-1em}
        %         &\includegraphics[trim={2cm 2cm 2cm 2cm},clip,width=.2\textwidth]{img/res/square_1_2_aniso_param.png} \\ %\hspace{-1em}
        %         \multicolumn{2}{c}{(anisotropic) integ. error: 0.00111369} & \\ 
        %     \end{tabular}
        % }
        \includegraphics[trim={2cm 2cm 2cm 2cm},clip,width=.2\textwidth]{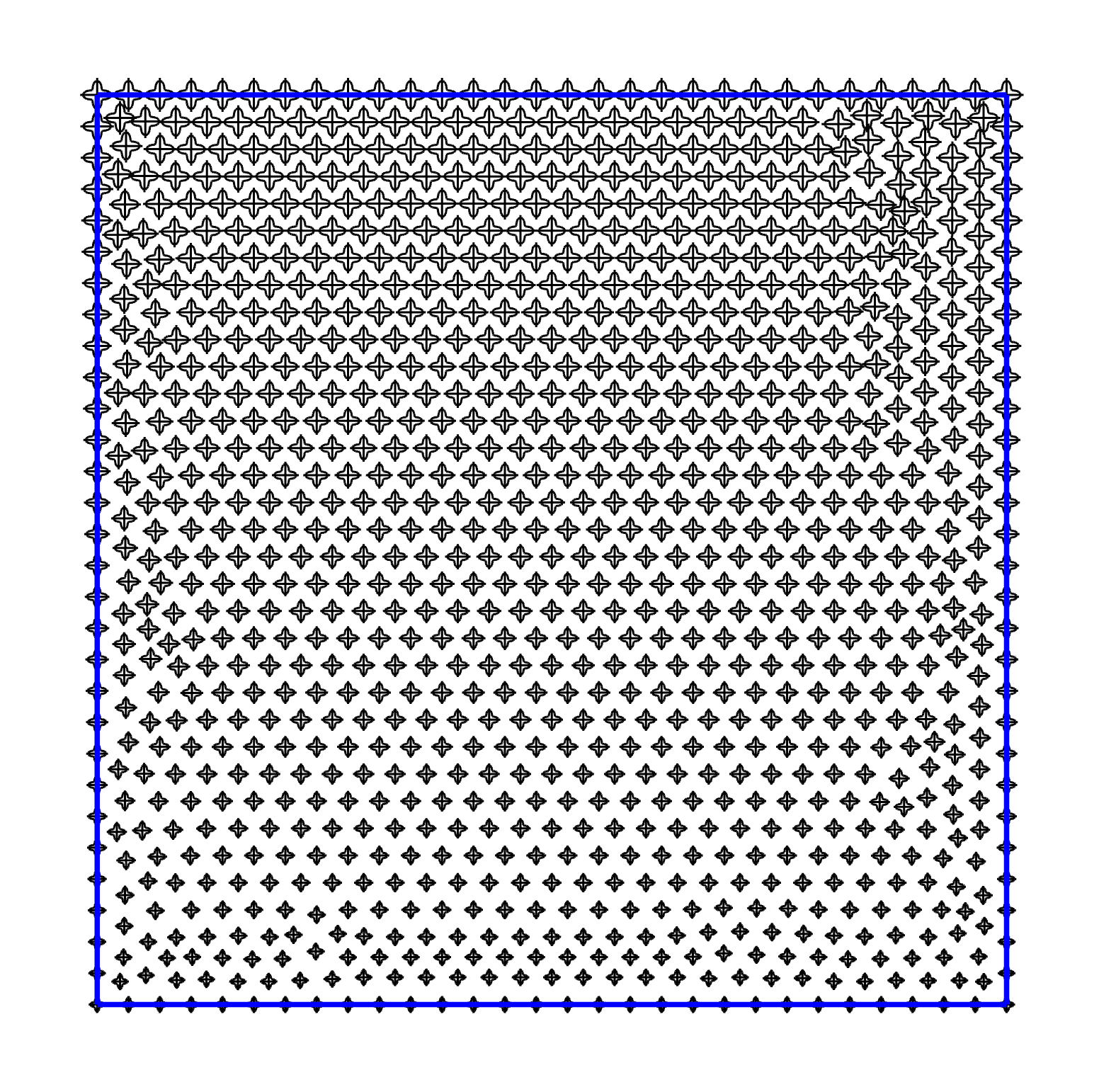} %\hspace{-1em}
        &\includegraphics[trim={2cm 2cm 2cm 2cm},clip,width=.2\textwidth]{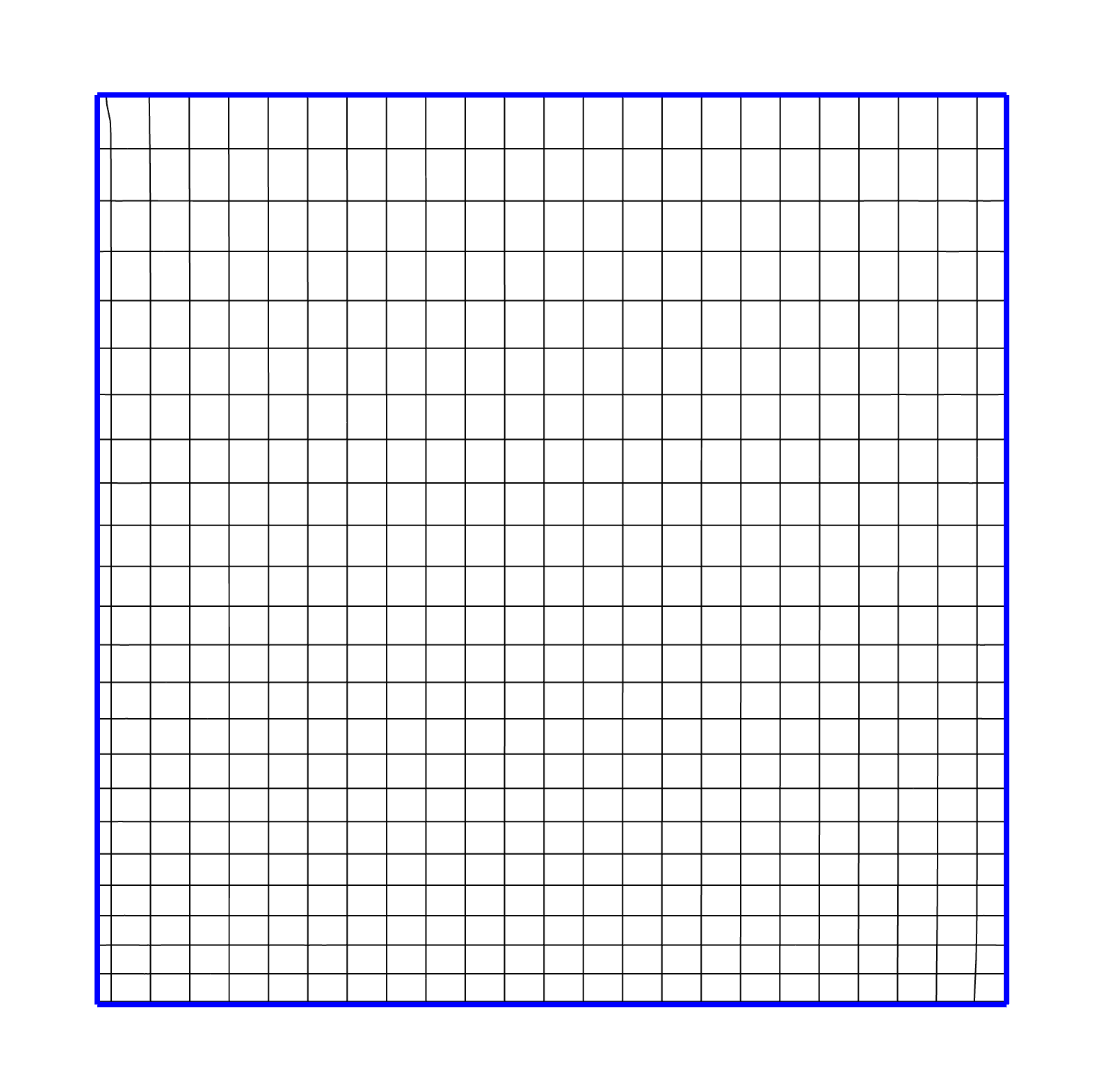} %\hspace{-1em}
        &\includegraphics[trim={2cm 2cm 2cm 2cm},clip,width=.2\textwidth]{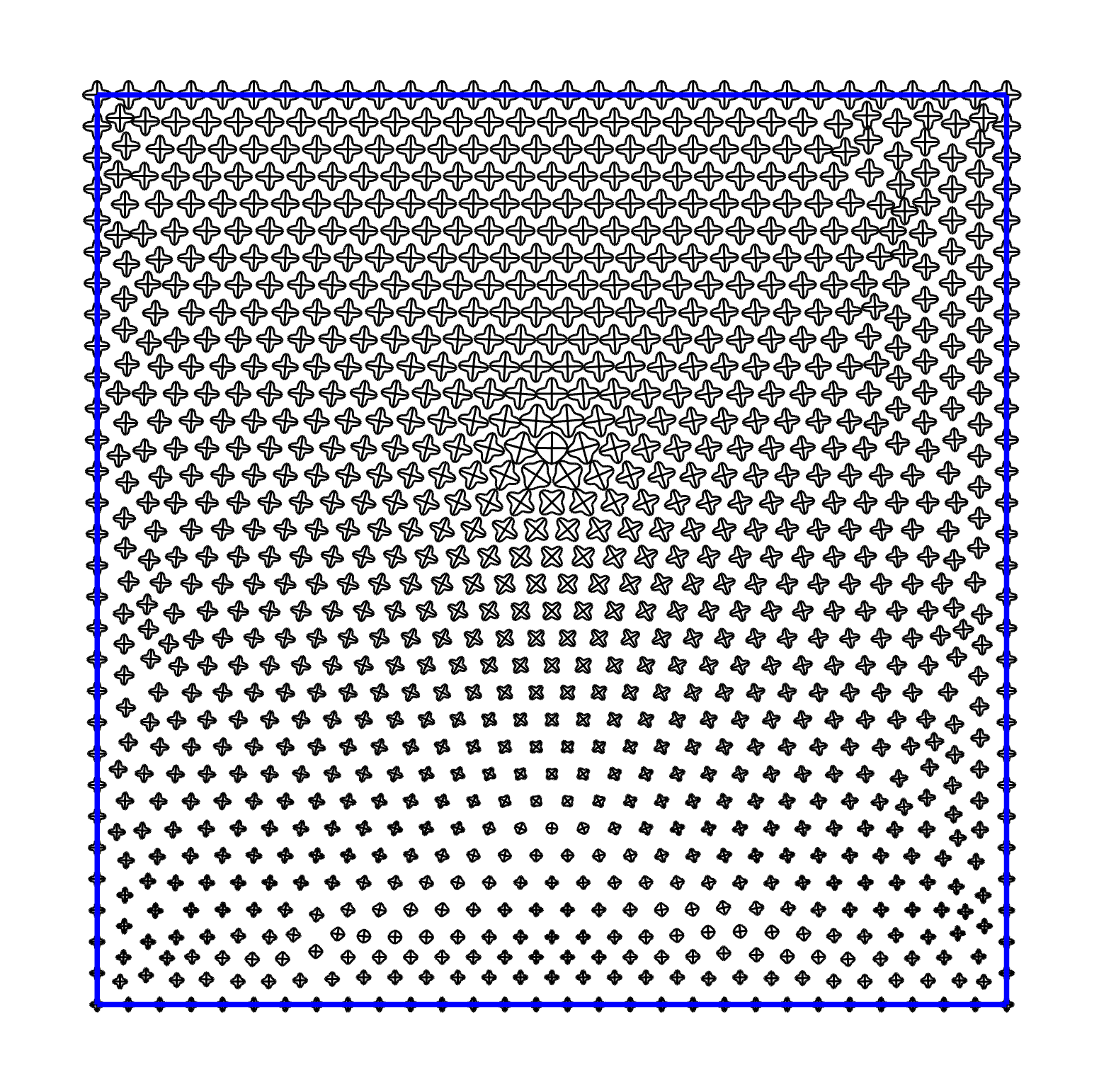} %\hspace{-1em}
        &\includegraphics[trim={2cm 2cm 2cm 2cm},clip,width=.2\textwidth]{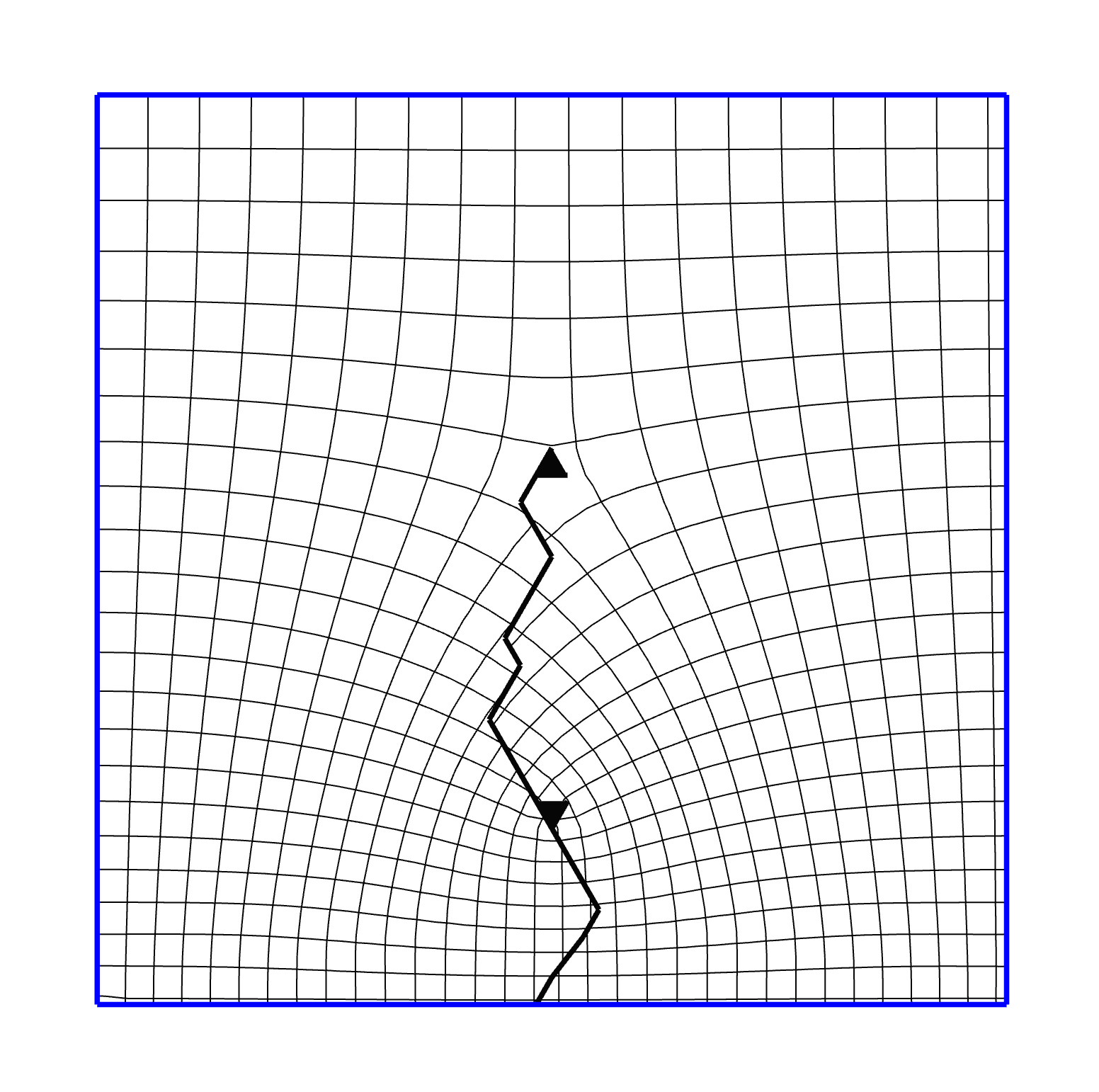} %\hspace{-1em}
        &\includegraphics[trim={2cm 2cm 2cm 2cm},clip,width=.2\textwidth]{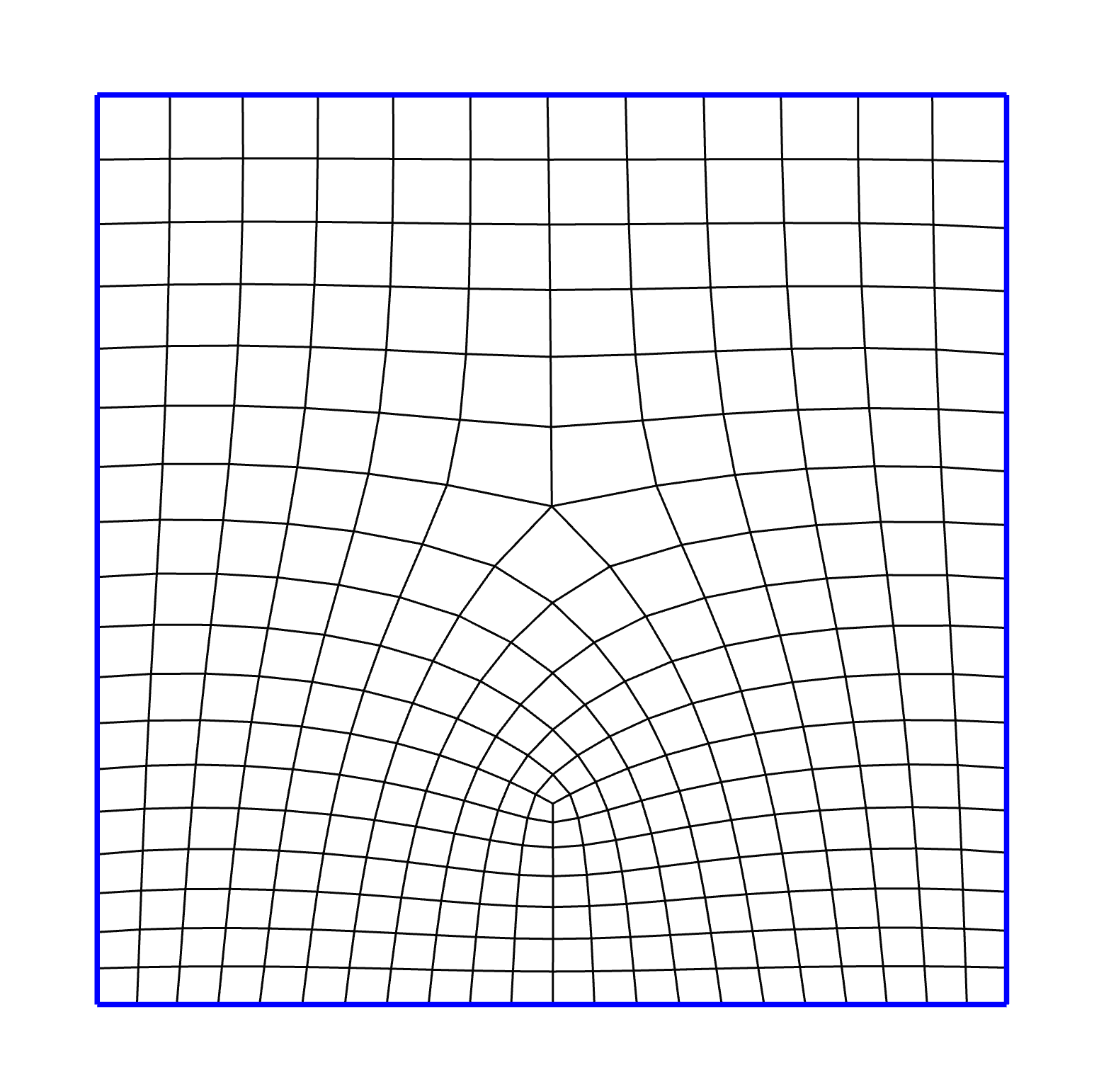} 
        \\
        \multicolumn{2}{c|}{integ. error: 0.0194898} & \multicolumn{2}{c}{integ. error: 0.00313874} &
        \\\hline
\addlinespace

        \multicolumn{5}{l}{(b) Size is 1 at bottom, 10 at top, and varies linearly on the sides.} \\
        \includegraphics[trim={2cm 2cm 2cm 2cm},clip,width=.2\textwidth]{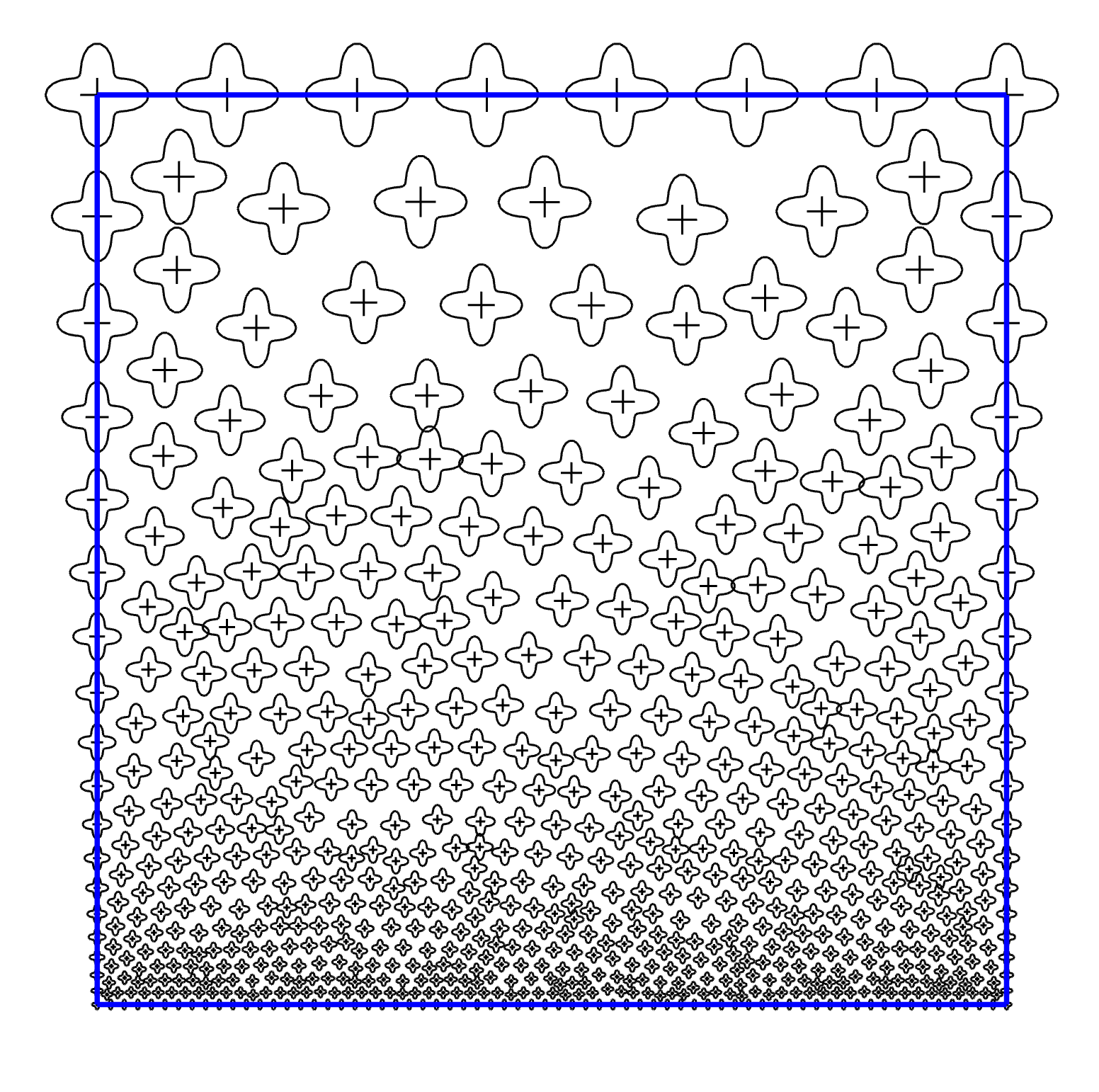} %\hspace{-1em}
        &\includegraphics[trim={2cm 2cm 2cm 2cm},clip,width=.2\textwidth]{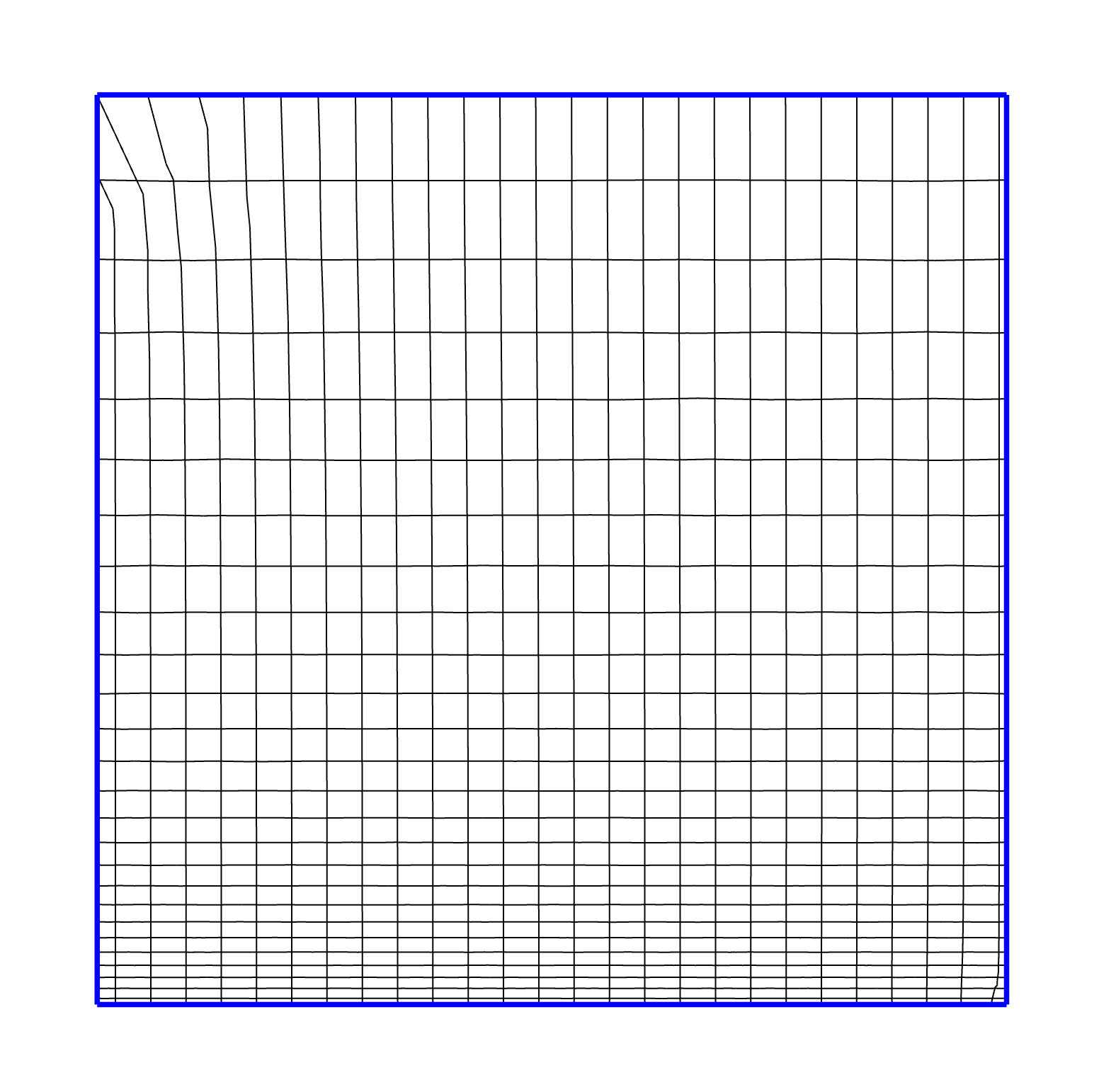} %\hspace{-1em}
        &\includegraphics[trim={2cm 2cm 2cm 2cm},clip,width=.2\textwidth]{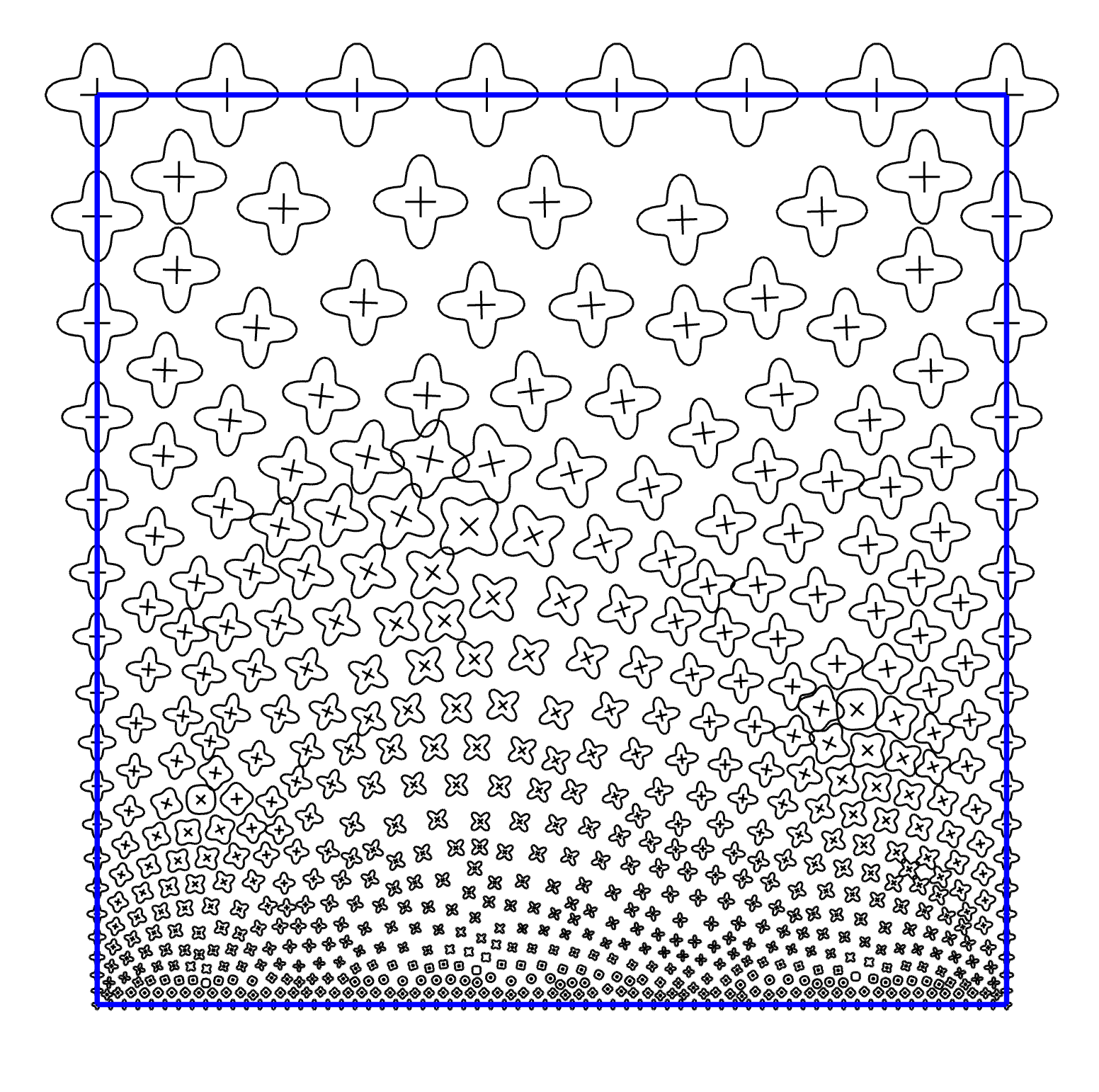} %\hspace{-1em}
        &\includegraphics[trim={2cm 2cm 2cm 2cm},clip,width=.2\textwidth]{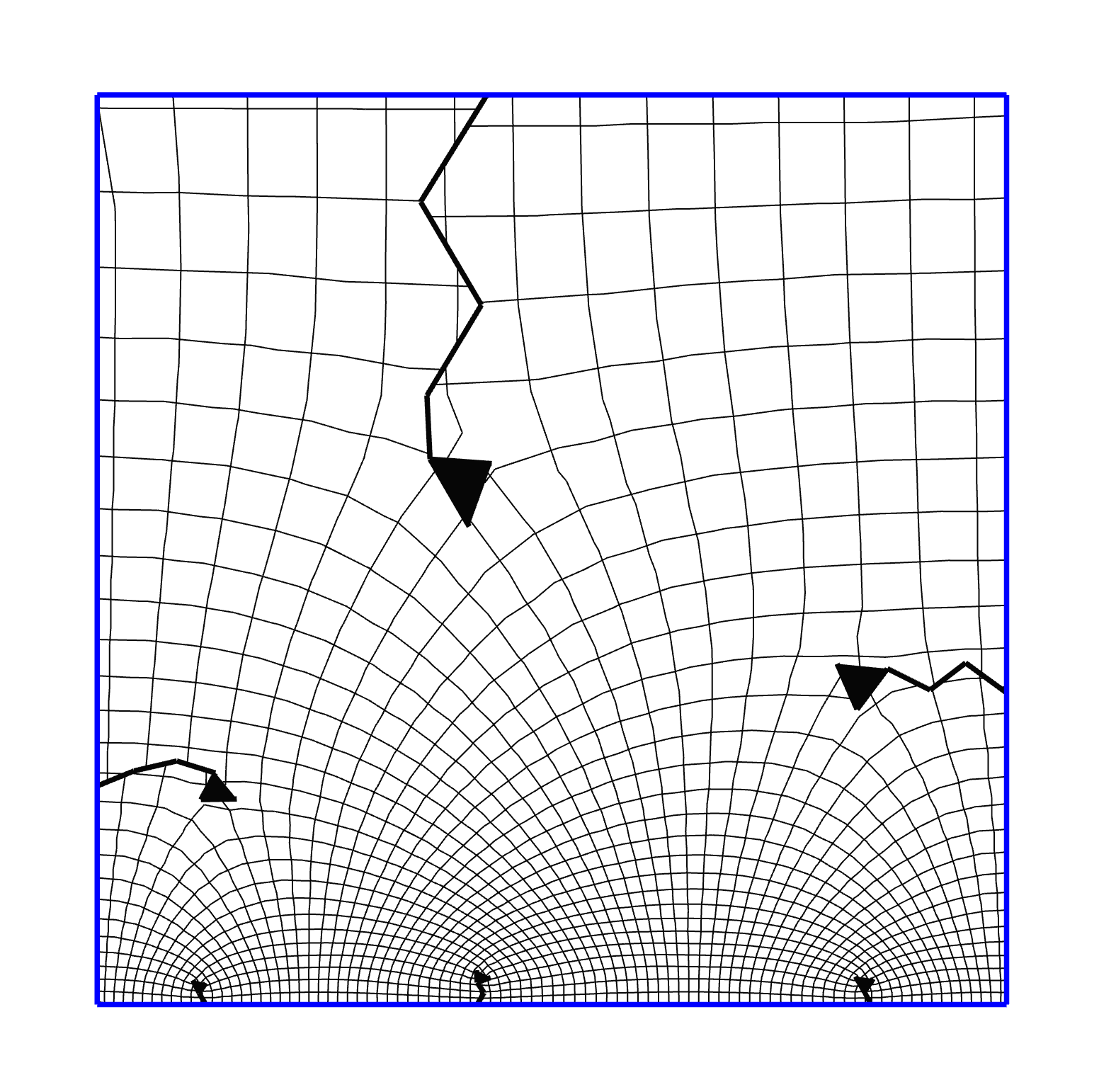} %\hspace{-1em}
        &\includegraphics[trim={2cm 2cm 2cm 2cm},clip,width=.2\textwidth]{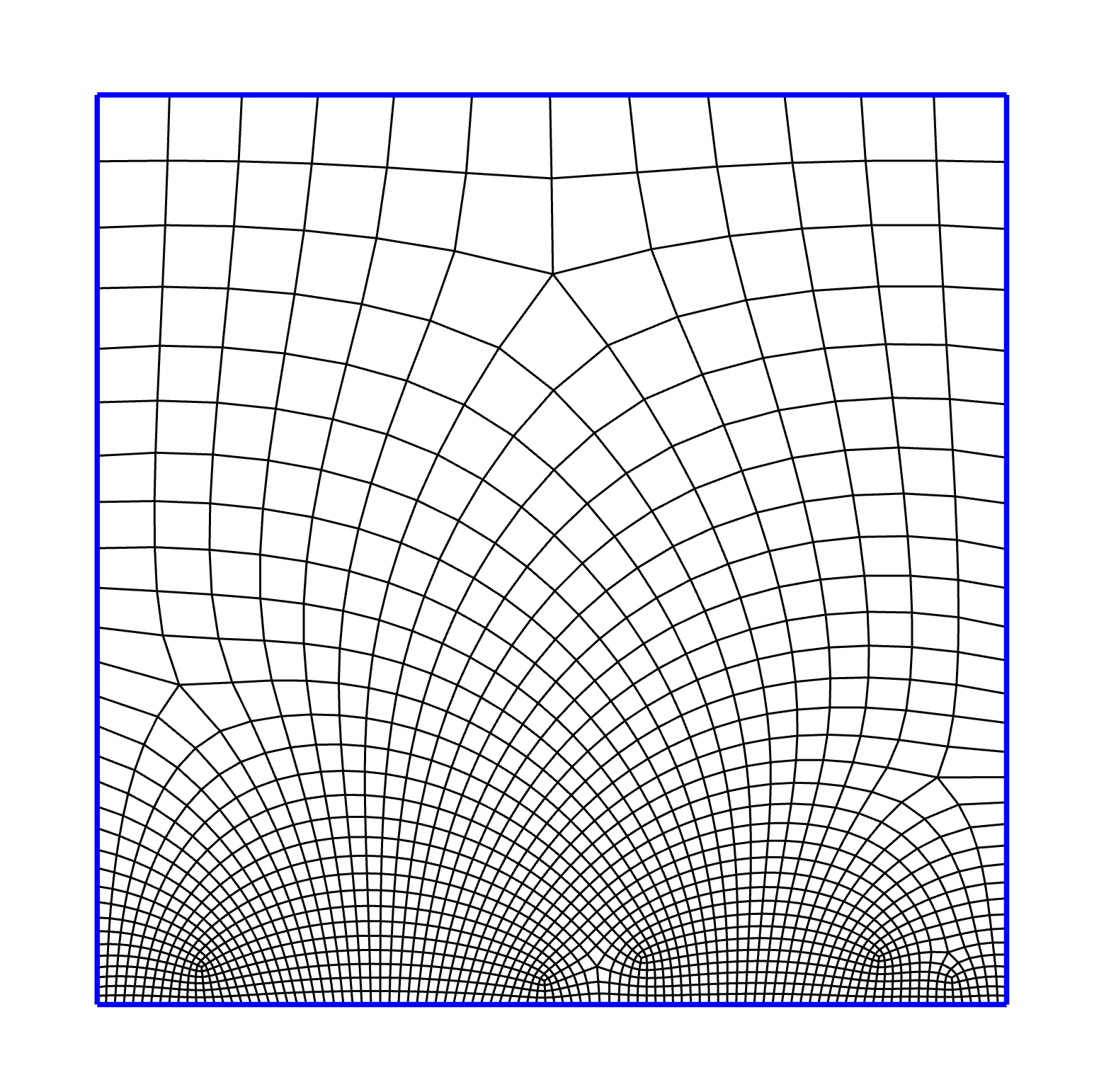} 
        \\
        \multicolumn{2}{c}{integ. error: 0.034417} & \multicolumn{2}{c}{integ. error: 0.00328246} & \\\hline \addlinespace
        \multicolumn{5}{l}{(c) Size is 4 on the inside and 1 on the outside. } \\
        \includegraphics[trim={15cm 30cm 15cm 0},clip,width=.19\textwidth]{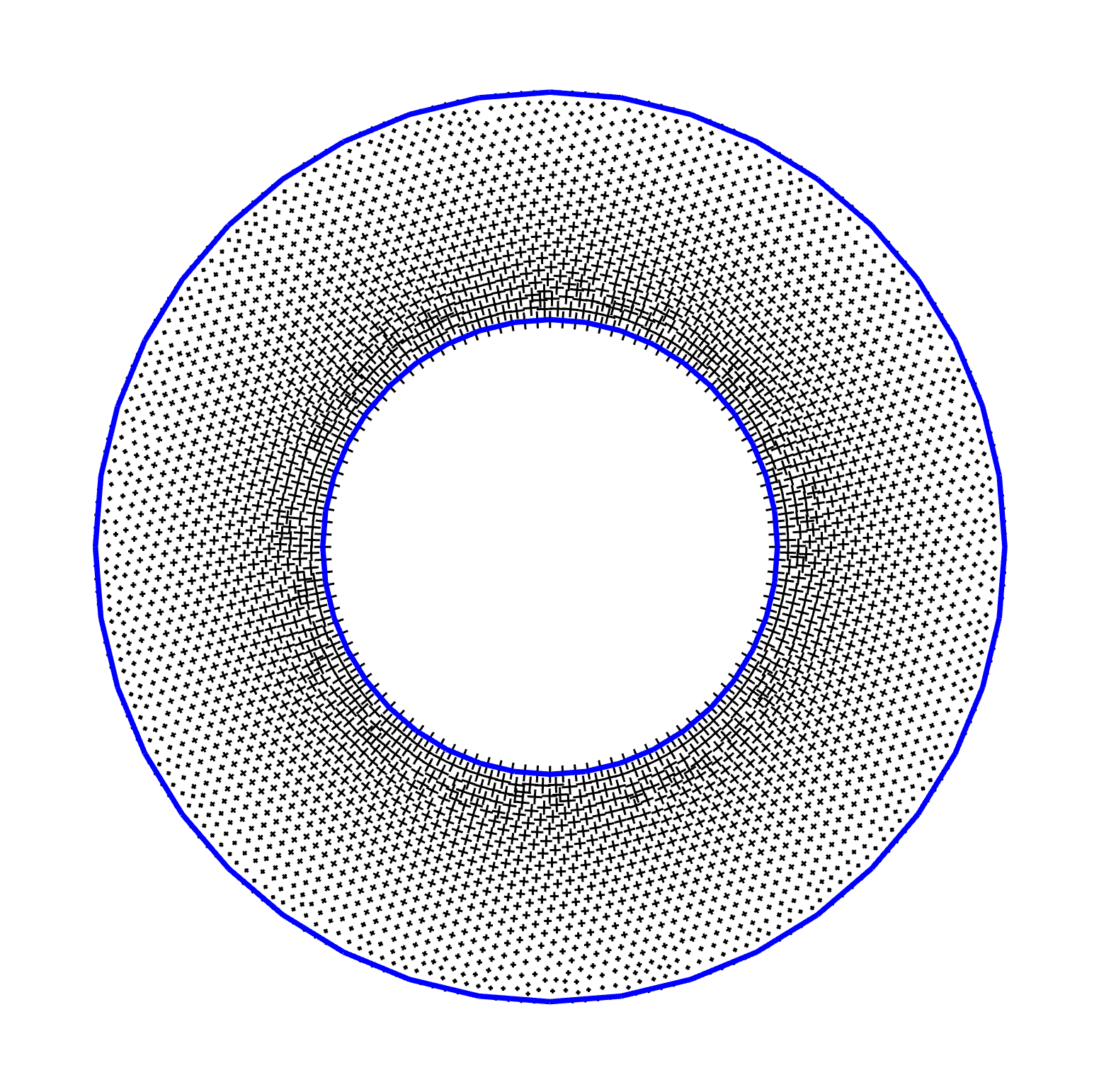} %\hspace{-1em}
        &\includegraphics[trim={15cm 30cm 15cm 0},clip,width=.19\textwidth]{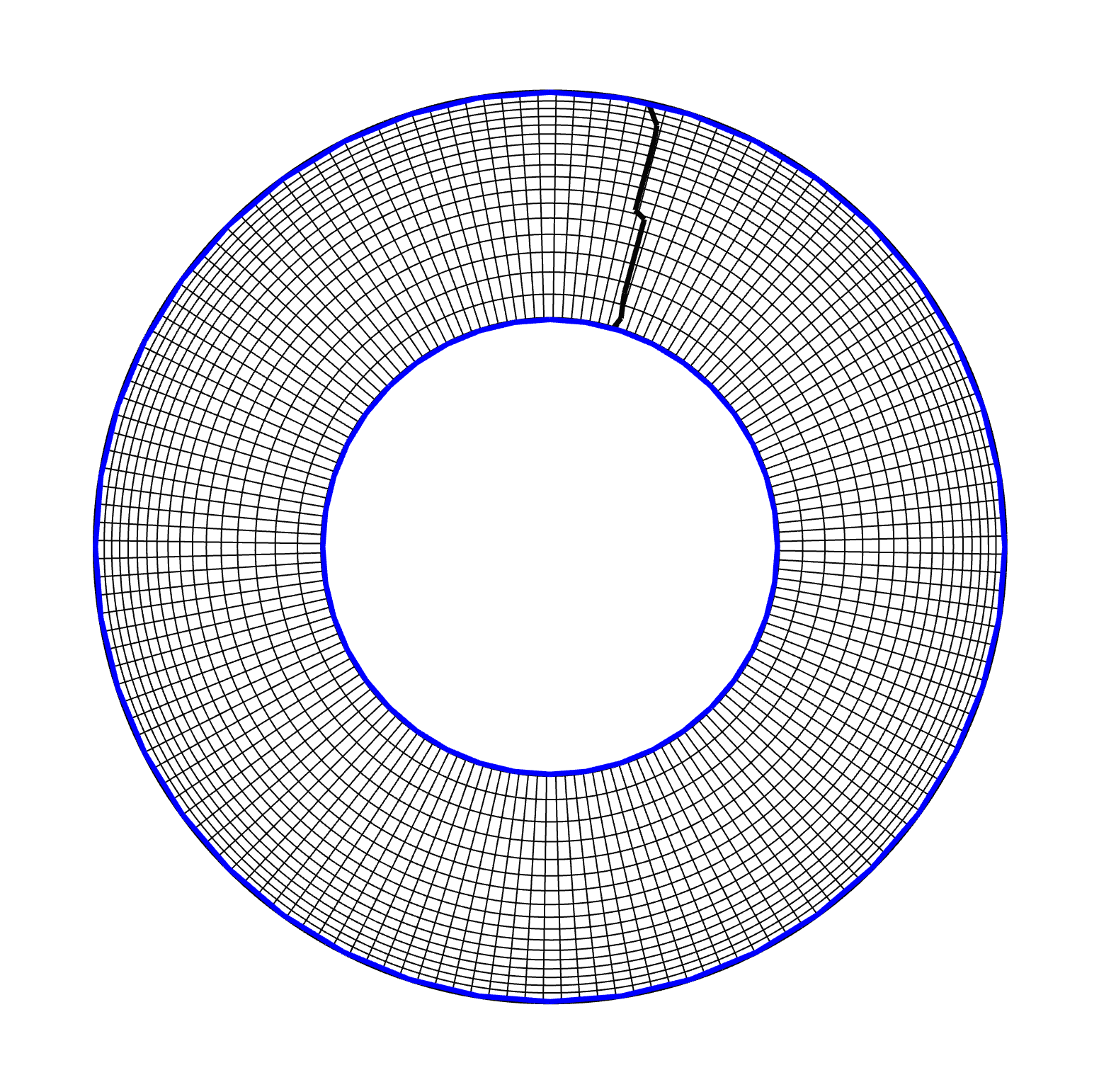} %\hspace{-1em}
        &\includegraphics[trim={15cm 30cm 15cm 0},clip,width=.19\textwidth]{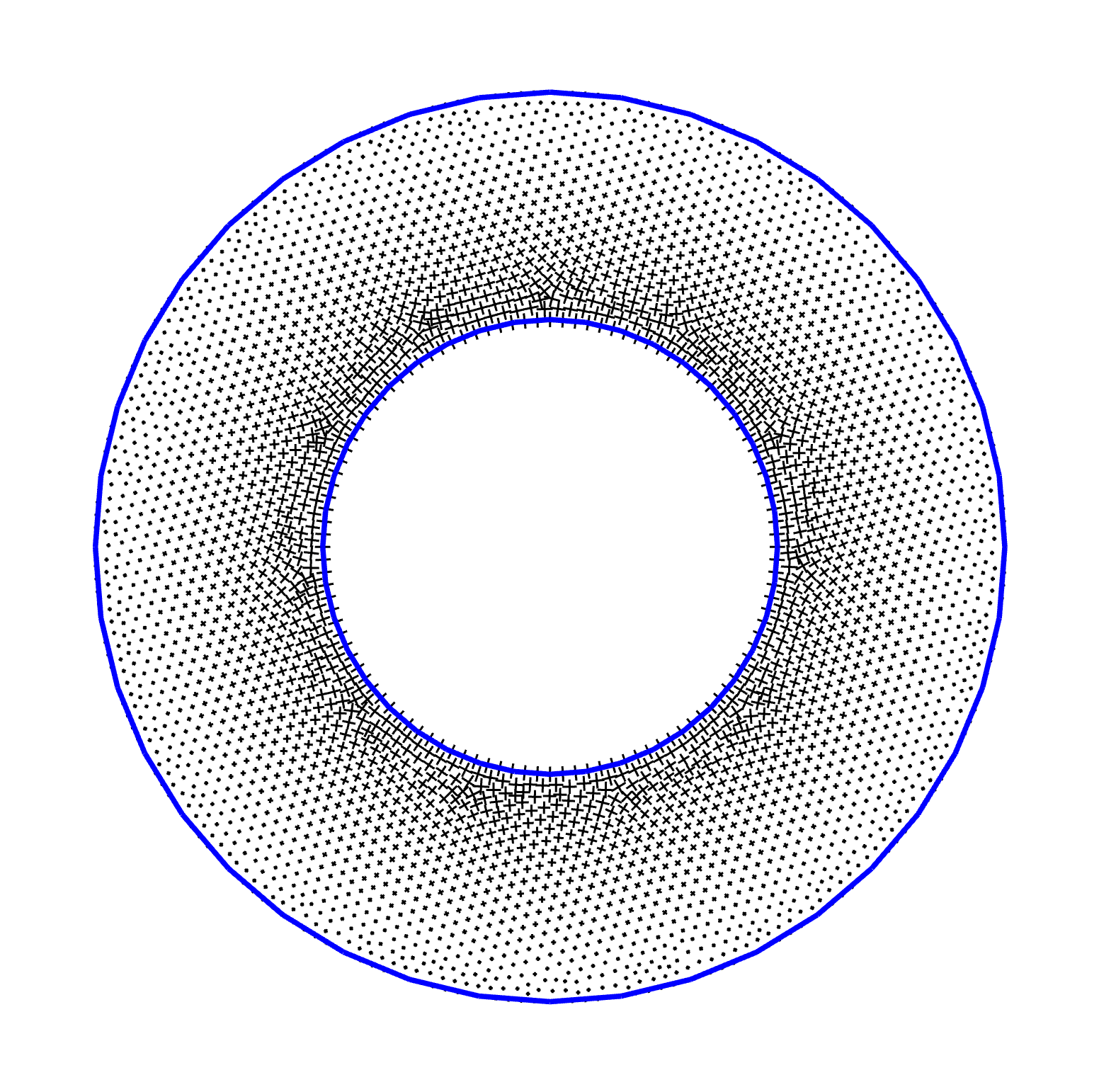} %\hspace{-1em}
        &\includegraphics[trim={15cm 30cm 15cm 0},clip,width=.19\textwidth]{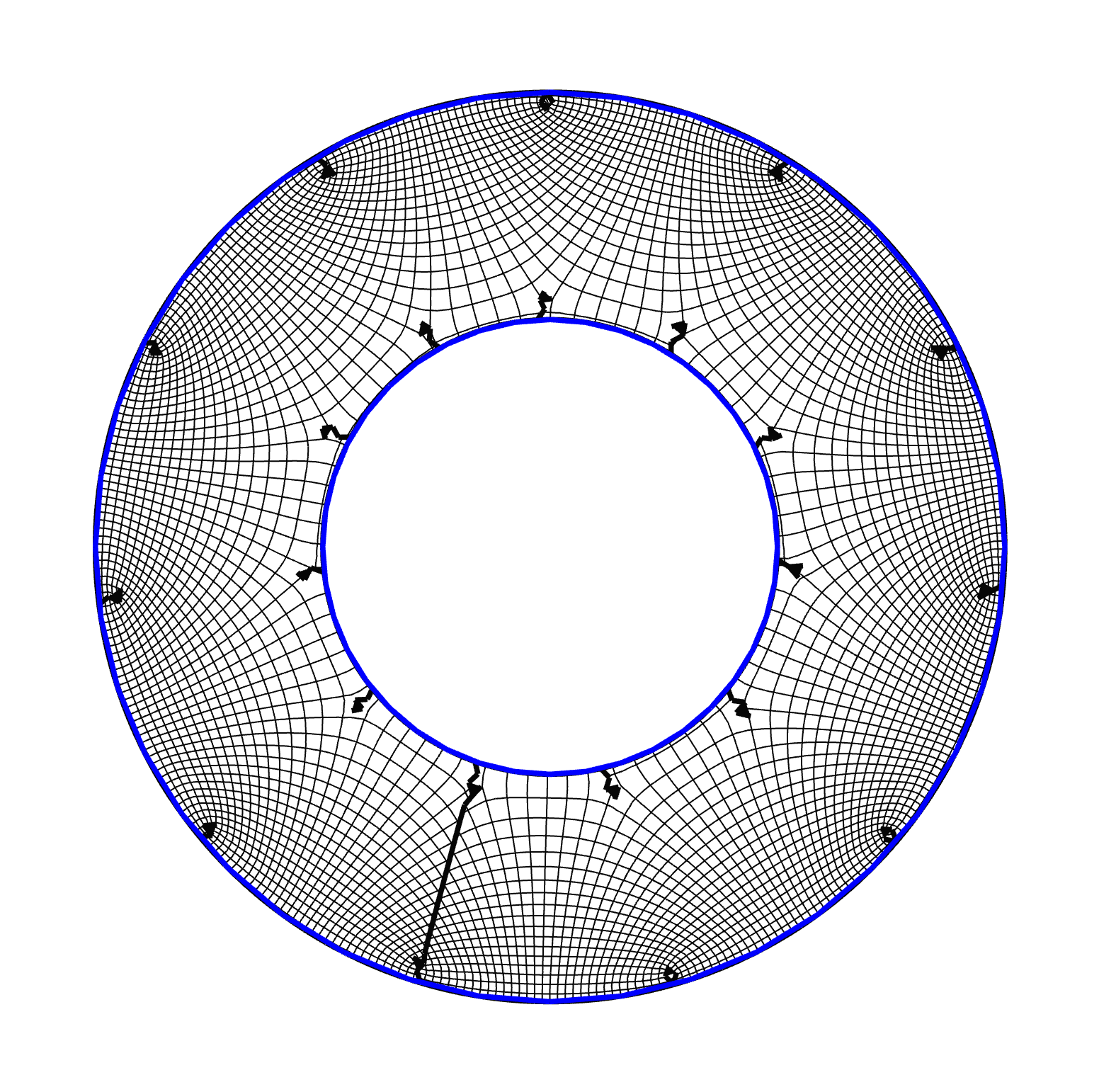}  %\hspace{-1em}
        &\includegraphics[trim={15cm 30cm 15cm 0},clip,width=.19\textwidth]{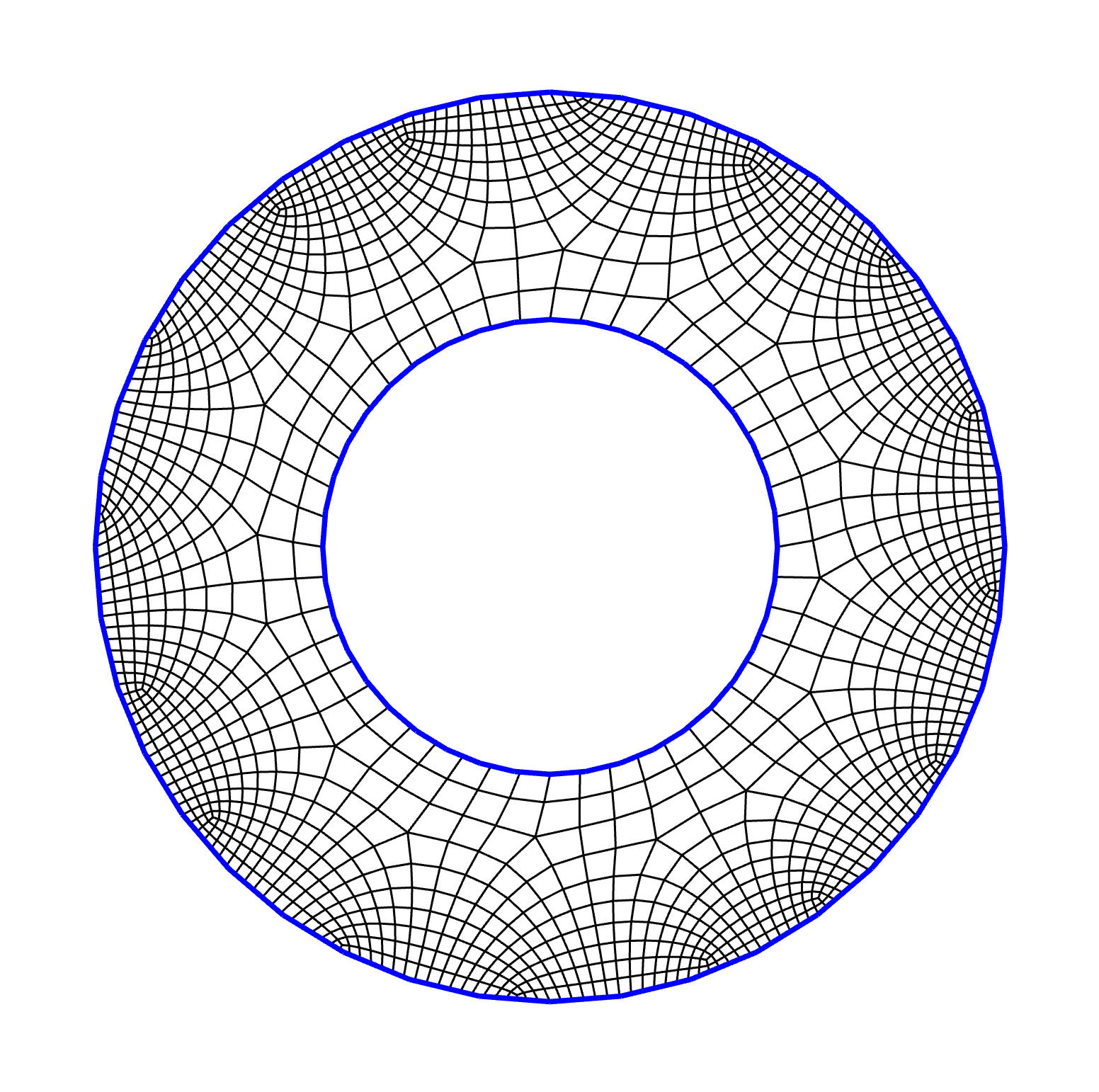}  %\hspace{-1em}
        \\
        \multicolumn{2}{c}{integ. error: 0.891866} & \multicolumn{2}{c}{integ. error: 0.100973} &
        \\\hline \addlinespace
        \multicolumn{5}{l}{(d) Size is 1 on the inside, 2 on the left outside and 1 on the left outside.} \\
        \includegraphics[trim={15cm 30cm 15cm 0},clip,width=.19\textwidth]{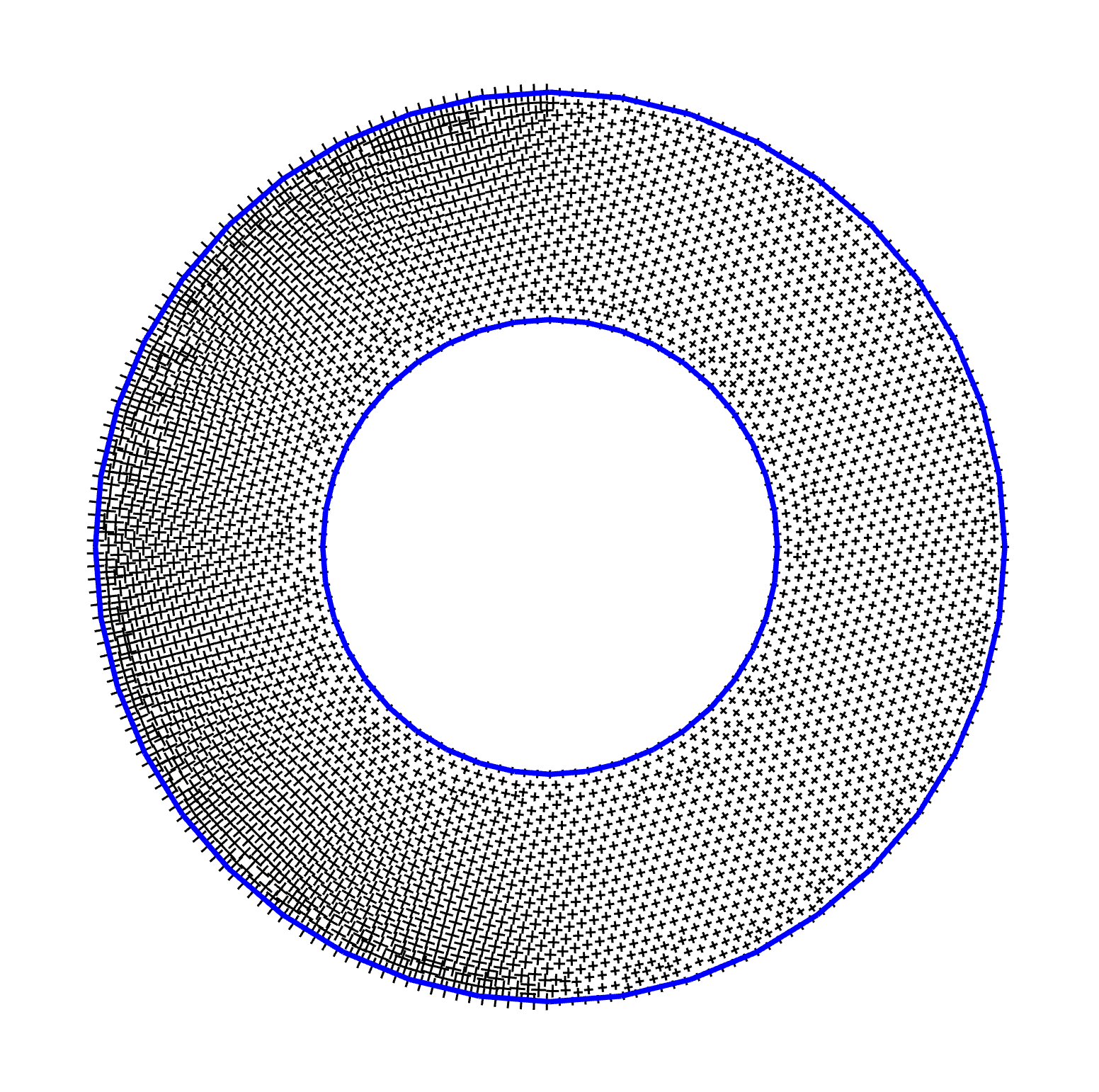} %\hspace{-1em}
        &\includegraphics[trim={15cm 30cm 15cm 0},clip,width=.19\textwidth]{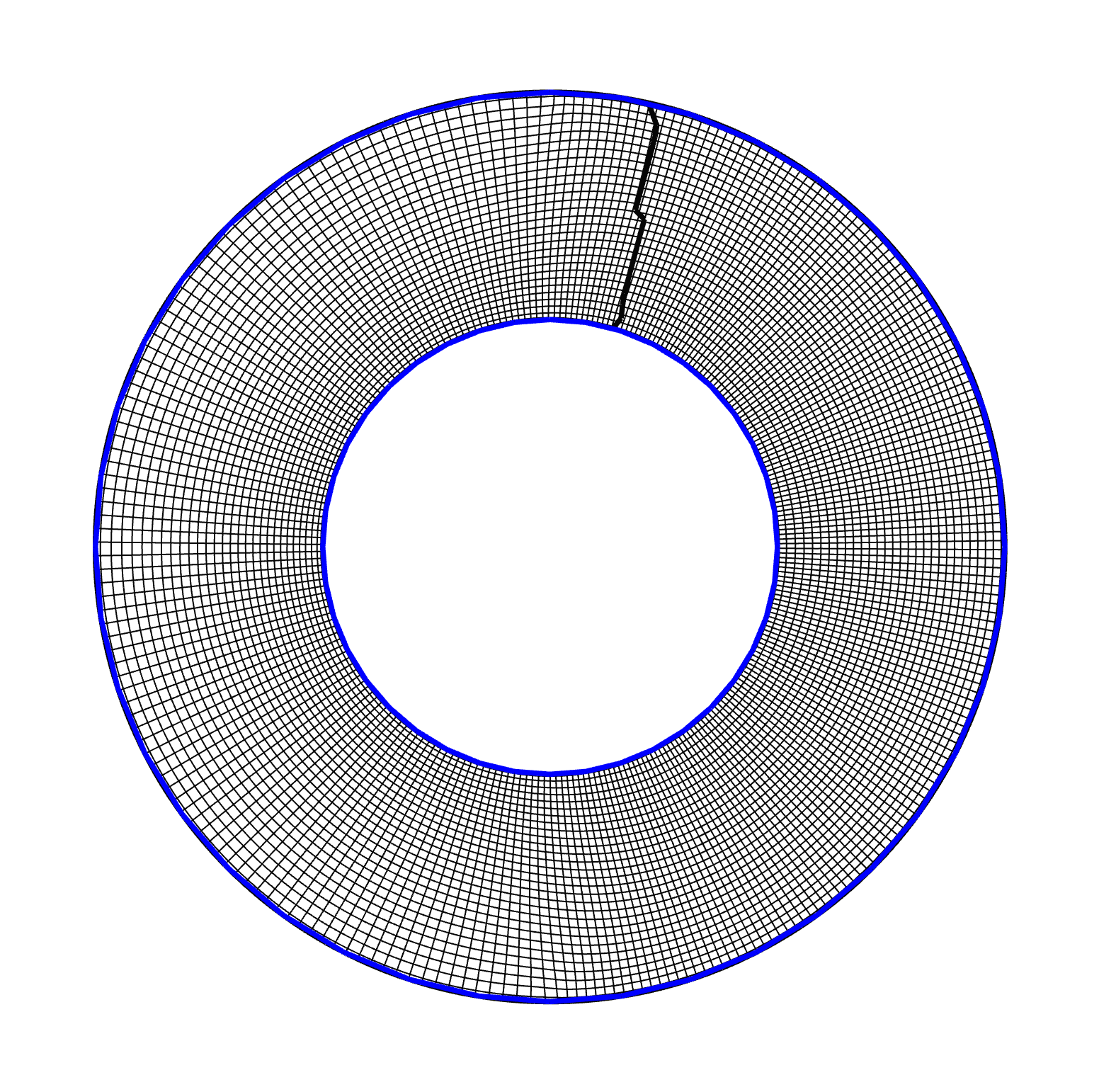} %\hspace{-1em}
        &\includegraphics[trim={15cm 30cm 15cm 0},clip,width=.19\textwidth]{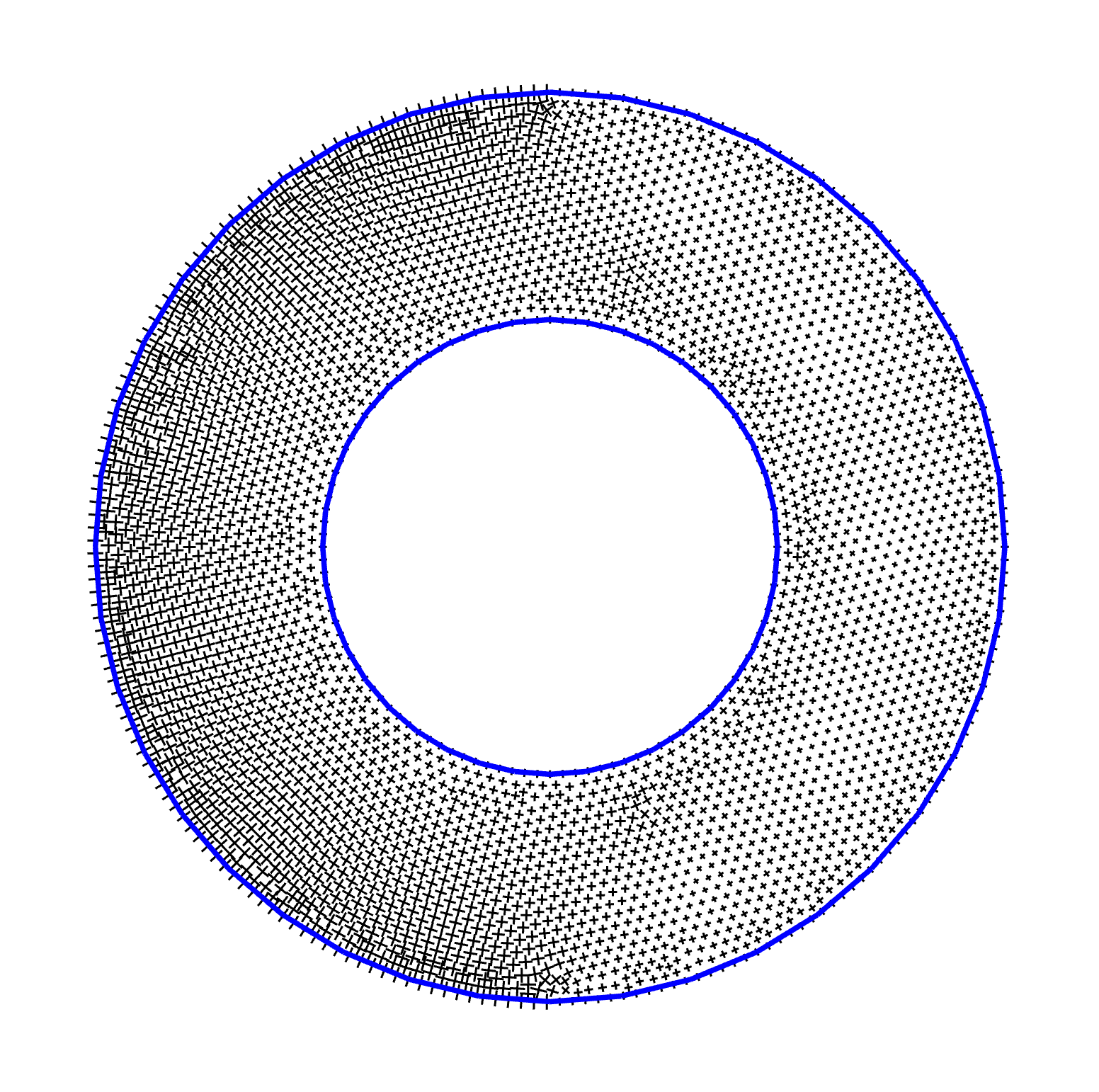} %\hspace{-1em}
        &\includegraphics[trim={15cm 30cm 15cm 0},clip,width=.19\textwidth]{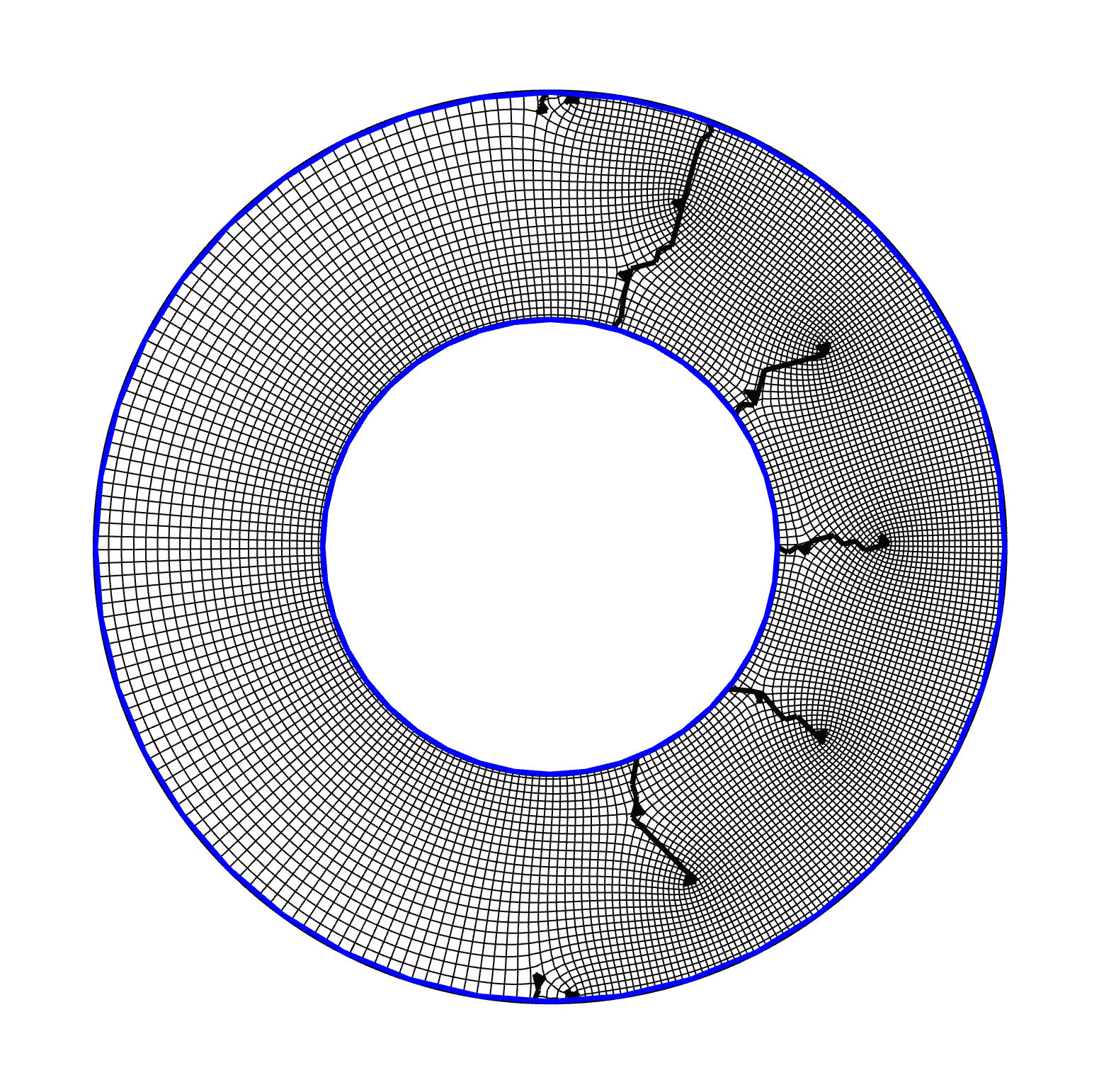} %\hspace{-1em}
        &\includegraphics[trim={15cm 30cm 15cm 0},clip,width=.19\textwidth]{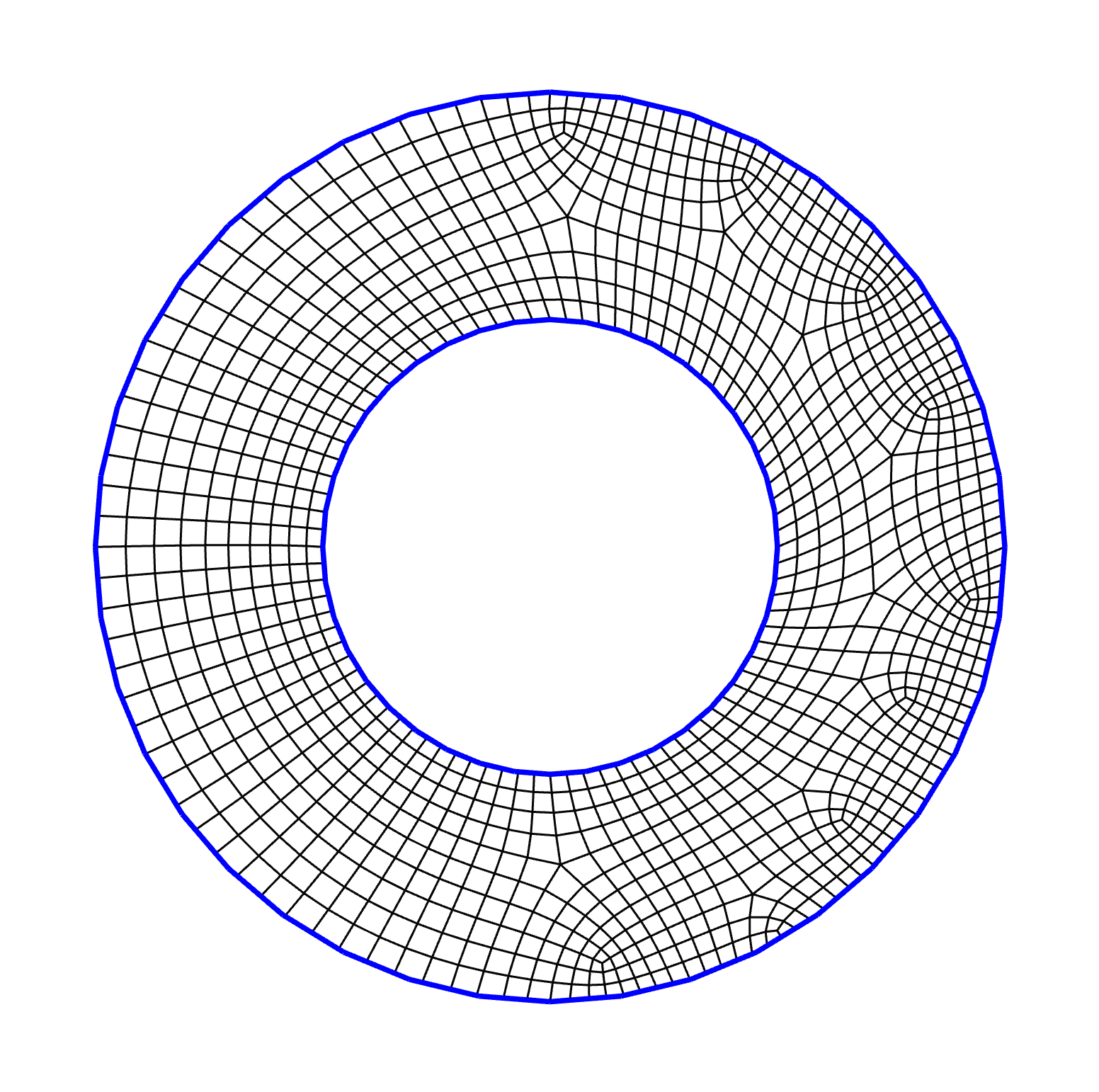} %\hspace{-1em}
        \\
        \multicolumn{2}{c}{integ. error: 0.172476} & \multicolumn{2}{c}{integ. error: 0.019616} &
        % (0.0194898/0.00313874 + 0.034417/0.00328246 + 0.891866/0.100973 + 0.172476/0.019616)/4 = 8.579973562
    \end{tabular}
    \caption{\update{Comparison of smooth frame fields and integrable frame fields
    when sizes are imposed on boundaries.
    For each frame field the corresponding parametrization is illustrated,
    and a quad mesh is computed using guidance from the integrable frame field.
    All frame fields are isotropic.
 %except for (a)
  %  where the anisotropic case is also illustrated.
    For cases (a) and (b) we also illustrate the polynomials $p_{\vb{T}}(\theta)$
    corresponding to the frames.
    % \note{TODO: Plots (c) and (d) are probably too small.
    % Maybe only show a part of it?}
    }}
    \label{fig:smooth_vs_integ}
\end{figure*}

\begin{figure}
    \begin{tabular}{@{}c@{}c}
        Isotropic frames & Anisotropic frames \\
        \includegraphics[trim={0cm 4cm 0cm 0cm},clip,width=.45\linewidth]{img/res/square_1_2_iso_param.png} &
        \includegraphics[trim={0cm 4cm 0cm 0cm},clip,width=.45\linewidth]{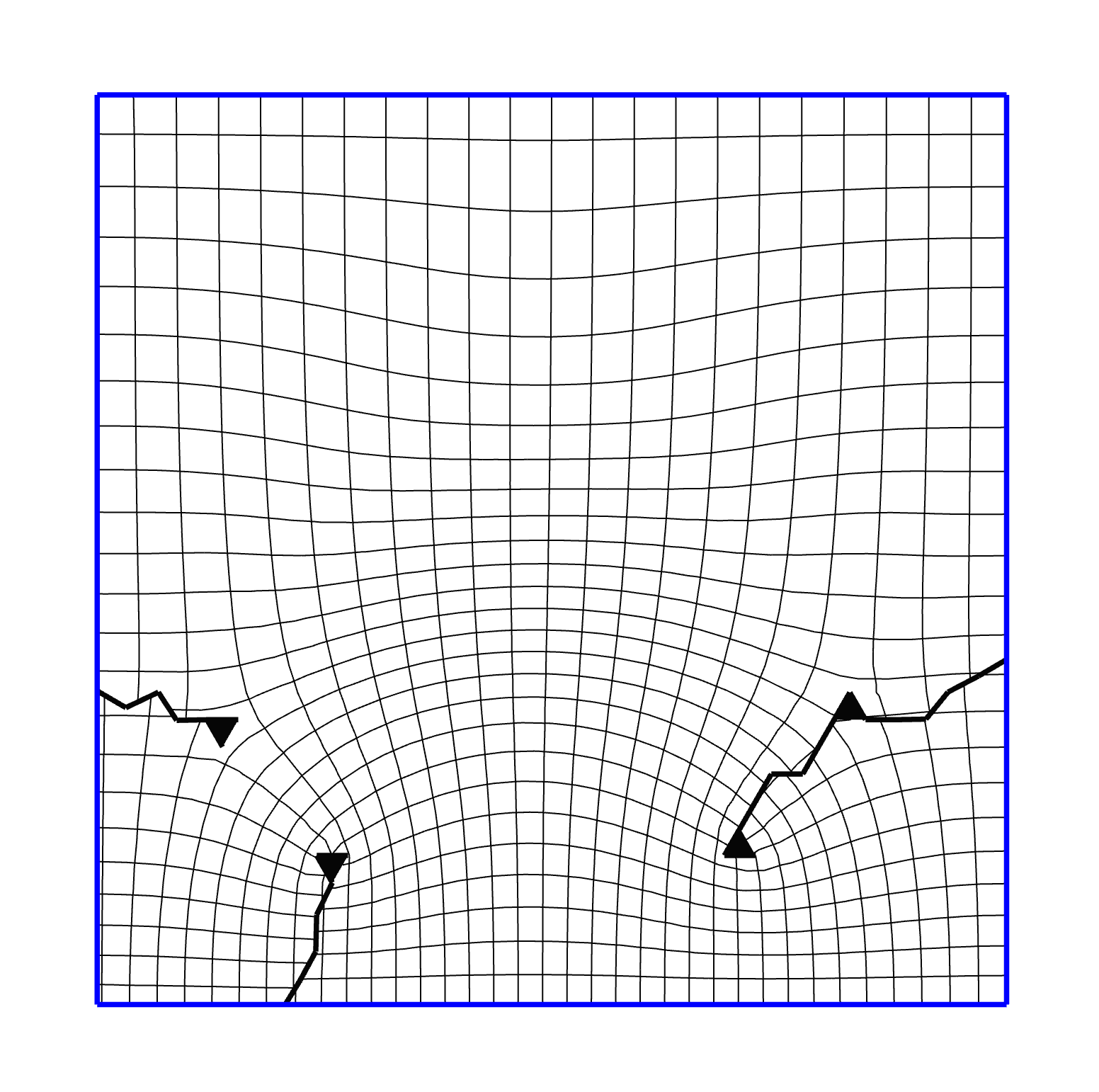} \\
        $e_\rm{integ} = 0.003$ & $e_\rm{integ} = 0.001$ \\
        \includegraphics[trim={0cm 3cm 0cm 0cm},clip,width=.43\linewidth]{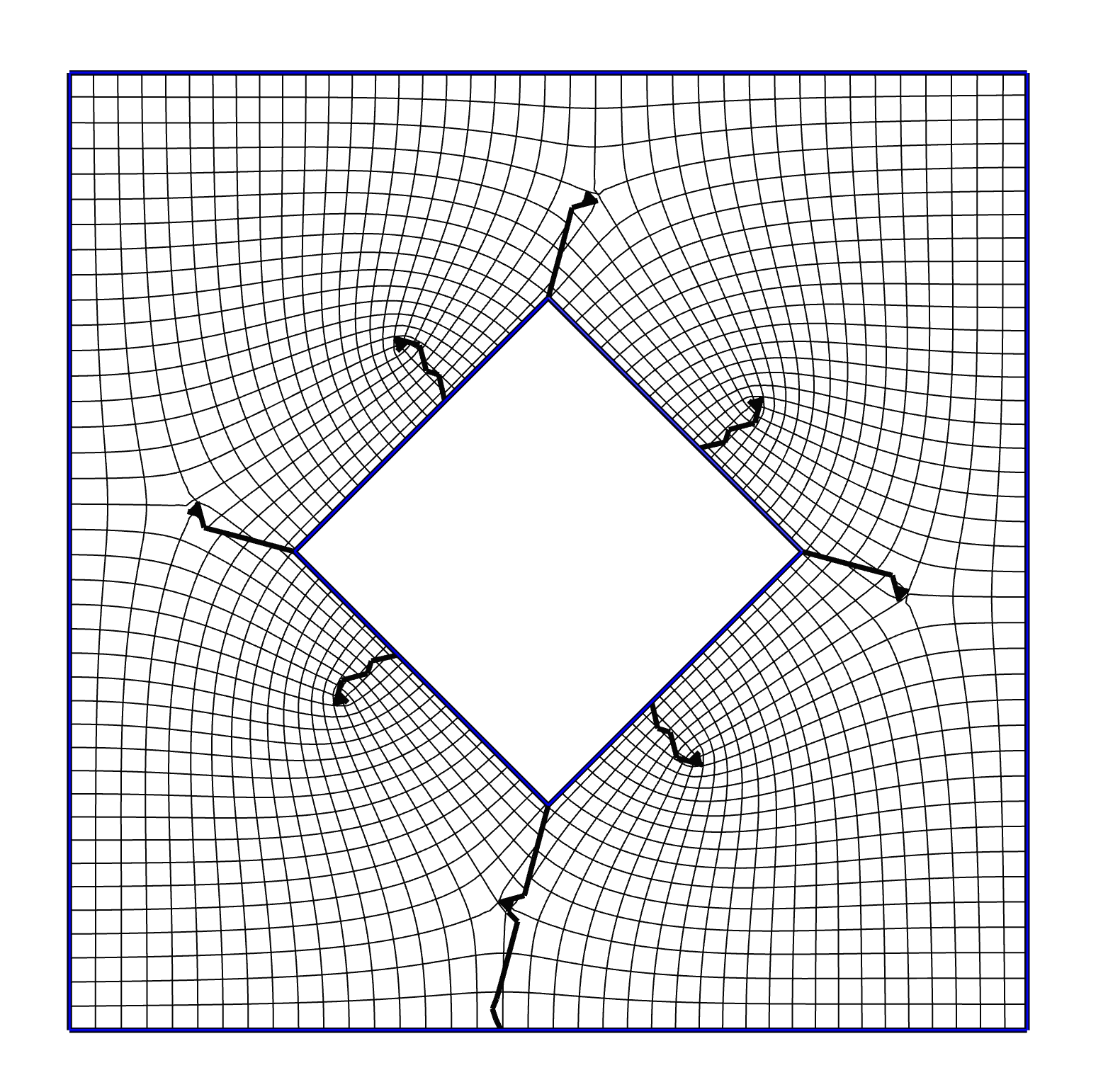} &
        \includegraphics[trim={0cm 3cm 0cm 0cm},clip,width=.43\linewidth]{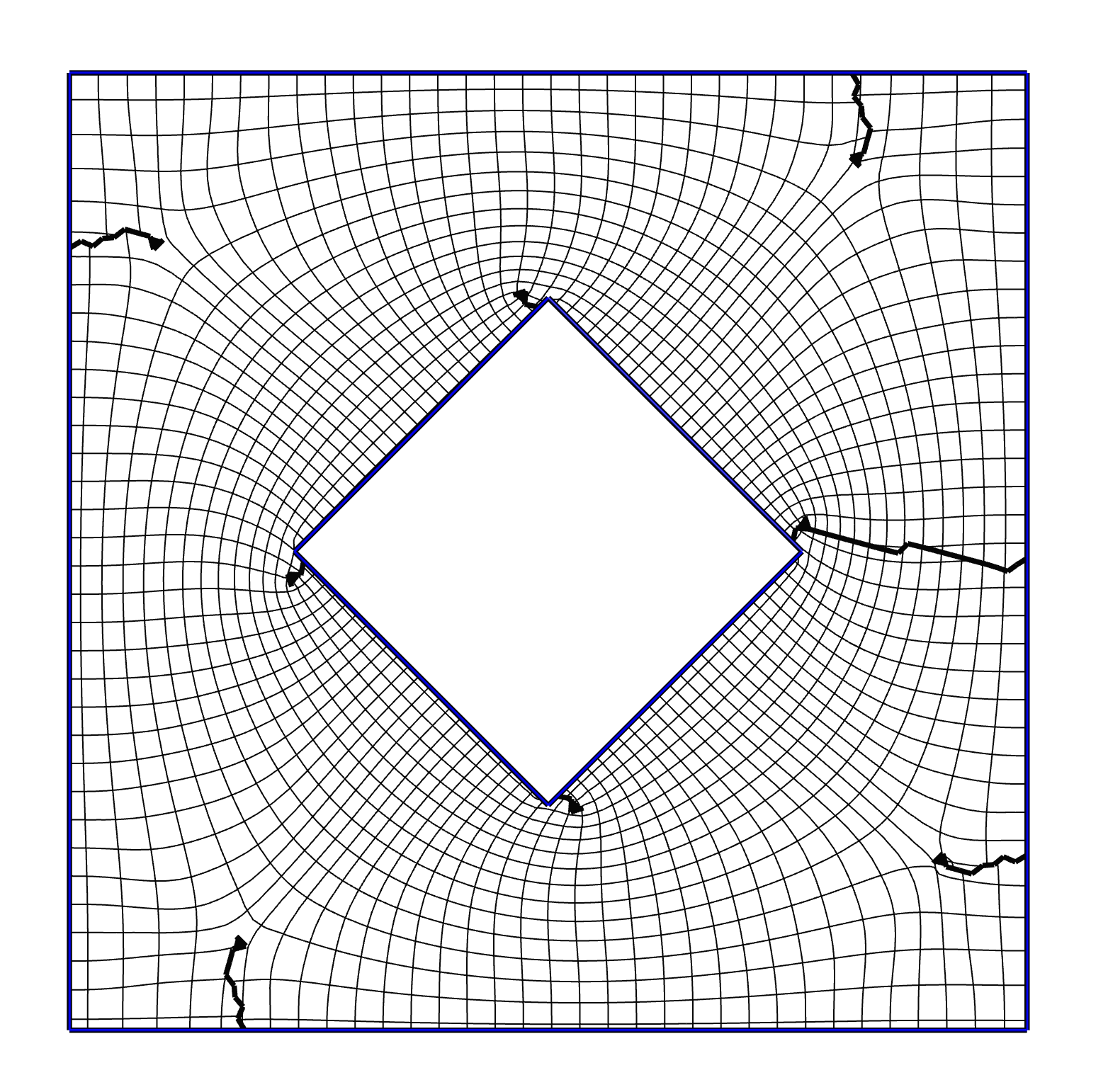} \\
        $e_\rm{integ} = 0.016$ & $e_\rm{integ} = 0.002$ 
        \\
        \includegraphics[trim={0cm 3.5cm 0cm 0cm},clip,width=.495\linewidth]{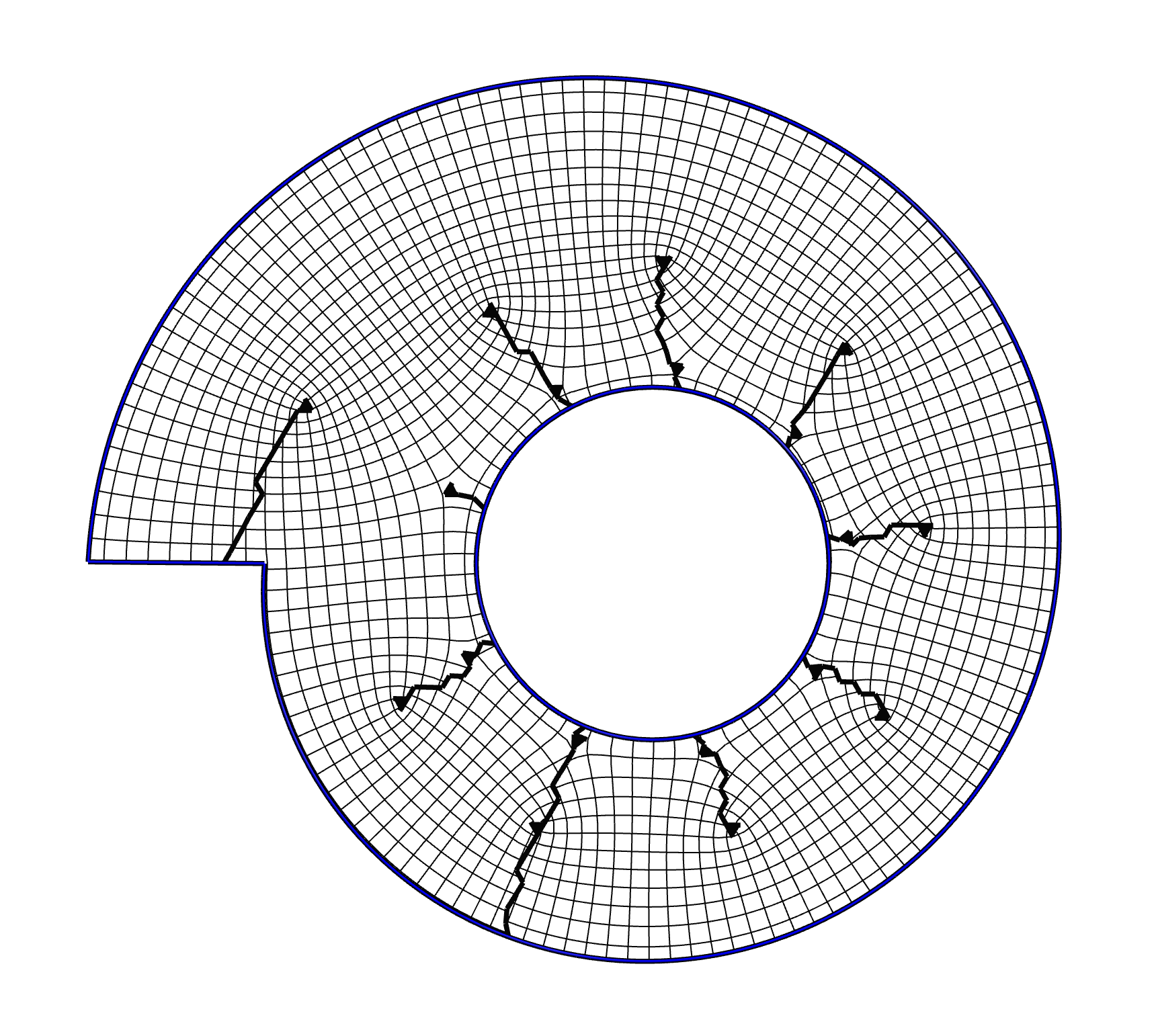} &
        \includegraphics[trim={0cm 3.5cm 0cm 0cm},clip,width=.495\linewidth]{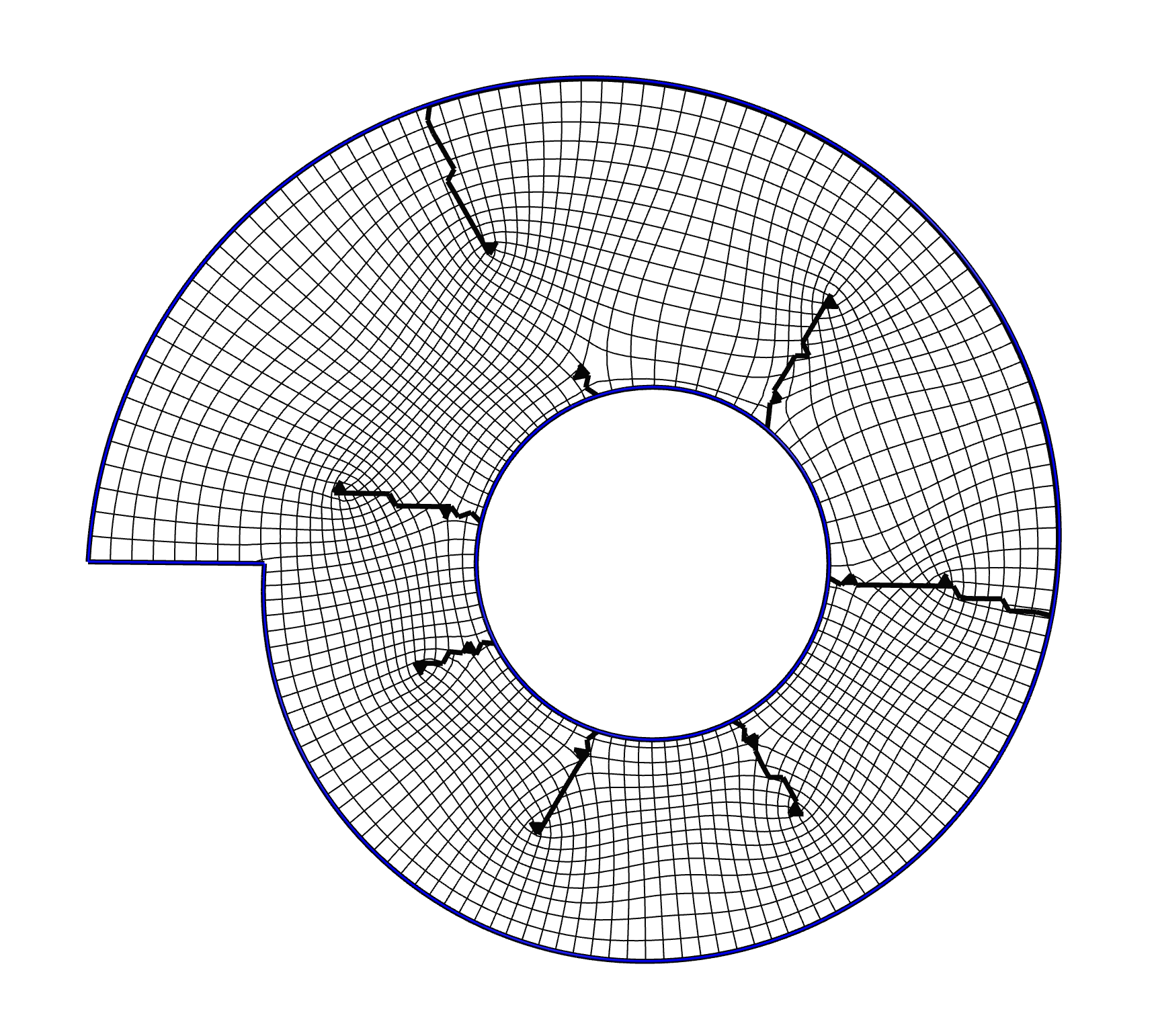} \\
        $e_\rm{integ} = 0.035$ & $e_\rm{integ} = 0.009$ \\
        
    \end{tabular}
    % \begin{subfigure}{.49\linewidth}
    %     \includegraphics[width=\linewidth]{img/res/tilted_squares_iso_param.png}
    %     \caption{Isotropic frames\\($e_\rm{integ} = 0.016$)}
    % \end{subfigure}
    % \begin{subfigure}{.49\linewidth}
    %     \centering
    %     \includegraphics[width=\linewidth]{img/res/tilted_squares_aniso_param.png}
    %     \caption{Anisotropic frames\\($e_\rm{integ} = 0.002$)}
    % \end{subfigure}
    \caption{\update{Seamless parametrizations computed for various models,
    including the \emph{nautilus} (bottom)
    in the isotropic and anisotropic settings.
    $e_\rm{integ}$ denotes the integration error.
    Boundary size conditions are \\
    (top) 1 at bottom and 2 at top, \\
    (middle) 1 on inner boundary and 2 on outer boundary, \\
    (bottom) 1 on inner and outer boundaries.
    }}
    \label{fig:aniso}
\end{figure}

% Aniso integrability error: 0.00111369

\paragraph{Comparison.}
The results of this comparison are illustrated in~\autoref{fig:smooth_vs_integ}.
On average, the integrable frame field achieves an integration error
\num{8.6} times smaller than the smooth frame field.
The smooth frame fields do not have the correct singularity configuration
to perform the required size transition.
This causes the corresponding parametrizations to be very anisotropic,
even though the frames are fully isotropic.
On the other hand, the integrable frame fields
have a set of singularities that ensure the size transitions,
and this is reflected in the parametrizations
where the isolines remain globally isotropic
and have the correct sizes at the boundaries.
We also see on testcase (d) that the solver is able to represent abrupt
size transitions through a tight pair of singularities.

\paragraph{Singularities.}
These results also exhibit how the field of tensors behaves in the vicinity of singularities.
As expected, the tensors become non-odeco as they narrow the singularities;
this allows the solver to represent singularities in a way
that is not too costly in terms of integrability energy.
However it also brings a limitation:
as the tensors are not odeco,
the assumptions needed for the integrability energy fail,
causing the frame field to not be exactly integrable close to singularities.
In practice, one needs to tune parameter $\epsilon$
to achieve an appropriate trade-off between singularity cost and integrability.

\paragraph{Isotropic vs anisotropic frames.}
\update{~\autoref{fig:aniso} compares the isotropic and anisotropic frame fields
obtained with our solvers on additional examples.
% The \emph{nautilus} model is notoriously challenging for frame field-driven quad meshing
% due to the presence of a limit cycle in fields produced by conventional solvers.
% Our method produces a singularity configuration that is free of limit cycles,
% and a valid quad mesh can be extracted.
The anisotropic solver produces a very distinct singularity configuration
compared to the isotropic solver, and achieves a much smaller integration error
thanks to the additional degrees of freedom.}

\paragraph{Limit cycles.}
\update{Conventional frame field-driven quad meshing methods
fail due to the presence of \emph{limit cycles} in the frame field,
which are the consequence of a non-meshable topology.
A standard test case where a limit cycle appears is the \emph{nautilus}
model, for which we show our results on~\autoref{fig:aniso}, bottom.
Our method successfully produces singularity configurations the are free of limit cycles,
and a valid quad mesh can be extracted.}

\paragraph{Existence of integrable frame fields.}
Given a set of boundary conditions,
an exactly integrable frame field does not exist in general;
in our case, by imposing sizes strongly on the boundaries,
an integrable isotropic frame field very often does not exist.
We refer to~\cite{jezdimirovic22} and the Abel-Jacobi framework
for a complete treatment of these aspects.
The non-existence of an integrable frame field
can explain why a Lie bracket of zero is not achievable.
\update{Moreover, the smaller integration error
achieved by allowing anisotropic frames (as illustrated in~\autoref{fig:aniso})
shows that the boundary conditions truly obstruct
the integrability. This can be alleviated
either by relaxing the boundary conditions or the isotropy constraint.}

\section{Conclusion and Future Work}
In this work, we have shown how to leverage
the theory of orthogonally decomposable tensors
to produce integrable frame fields.
We have studied the tensor eigenvalue perturbation problem
and found simple expressions for the eigenvalue sensitivity of tensors.
This result enables us to write a frame field's Lie bracket
(and thus, the integrability optimization problem)
solely in terms of its tensor representation.
The integrable frame field can be integrated 
to a seamless parametrization of which we have full control
over size and orientation through the frame field optimization;
this is a feature that is often overlooked in field-based meshing methods.
We believe this contribution is an important step towards
a robust, flexible (in terms of size and orientation prescriptions)
and optimal quadrilateral mesher.

\paragraph{Integrable non-odeco fields.}
An important limitation of our approach is that
the tensor field is assumed to be odeco
for the Lie bracket expression to be valid.
This causes the frame field to not be exactly integrable near singularities.
A possible solution for this is to remove the odeco assumption
when writing out the eigenvalue perturbation problem.
This makes it possible to write a Lie bracket
that remains valid even if the tensor field is non-odeco
(at least in the isotropic case).
The price to pay is that the integrability condition becomes
a more complex, rational expression of the tensor coefficients.
We leave the investigation of this approach for future work.

\paragraph{The 3D case.}
The major advantage of our methodology is that it offers a natural extension
to compute 3D integrable frame fields;
indeed, the theory of odeco tensors,
as well as the results on eigenvalue sensitivity,
remain valid in arbitrary dimensions.
The only obstacle remaining is to express
the integrability condition in terms of the algebraic frame representation
We are confident that a solution can be found in future work.

\section*{Acknowledgments}
Mattéo Couplet is a Fellow of the Belgian Fund for Scientific Research (F.R.S.-FNRS).
% This research is also supported by the European Research Council (project 

\newpage

\begin{appendices}

\section{Eigenvalue sensitivity for odeco tensors} \label{app:sensitivity}
Let $\vb{T}$ be a fully symmetric fourth-order orthogonally decomposable tensor
of dimension $n$, i.e., $\vb{T} = \sum_{i=1}^n \lambda_i \vb{\hat{v}}_i^4$,
or, in index notation, 
\begin{equation}
    T_{j_1j_2j_3j_4} = \lambda_i \hat{v}_{i,j_1} \hat{v}_{i,j_2} \hat{v}_{i,j_3} \hat{v}_{i,j_4},
\end{equation}
with orthogonal unit vectors $\vb{\hat{v}}_i$.
Consider the contraction of $\vb{T}$ with one of its eigenvectors, $\vb{\hat{v}}_k$:
\begin{equation}
    \begin{aligned}
        T_{j_1j_2j_3j_4} \hat{v}_{k,j_4} &= \lambda_i \hat{v}_{i,j_1} \hat{v}_{i,j_2} \hat{v}_{i,j_3} \hat{v}_{i,j_4} \hat{v}_{k,j_4}, \\
        &= \lambda_i \hat{v}_{i,j_1} \hat{v}_{i,j_2} \hat{v}_{i,j_3} \delta_{ik} \\
        &= \lambda_k \hat{v}_{k,j_1} \hat{v}_{k,j_2} \hat{v}_{k,j_3},
    \end{aligned}
\end{equation}
or, written compactly, $\vb{T}\vb{\hat{v}}_k = \lambda_k \vb{\hat{v}}_k^3$.
Contracting again and following the same procedure
we find that $\vb{T}\vb{\hat{v}}_k^2 = \lambda_k \vb{\hat{v}}_k^2$ and
$\vb{T}\vb{\hat{v}}_k^3 = \lambda_k \vb{\hat{v}}_k$,
the latter being the definition of a tensor eigenvector.
Consider now a specific eigenpair $(\lambda,\vb{w})$, with $\vb{w}$ having unit norm.
We wish to express the variation of the eigenpair $(\delta\lambda, \delta\vb{w})$
given some perturbation of the tensor $\delta\vb{T}$.
We write the remainder of the proof in compact notation for brevity
but the same steps can be performed in index notation.
Since the eigenvectors have unit norm, any eigenvector is orthogonal to its variation:
\begin{equation}
    (\vb{w}+\delta\vb{w})\cdot(\vb{w}+\delta\vb{w}) = 1 \ \Rightarrow \ \vb{w}\cdot\delta\vb{w} = 0,
\end{equation}
neglecting the higher-order terms $(\delta\vb{w}\cdot\delta\vb{w})$.
Writing the eigenvector definition $\vb{T}\vb{w}^3 = \lambda\vb{w}$
for the new eigenpair, we find
\begin{equation}
    \begin{aligned}
        (\vb{T}+\delta\vb{T})(\vb{w}+\delta\vb{w})^3 &= \lambda\vb{w} + \delta(\lambda\vb{w}) \\
        \cancel{\vb{T}\vb{w}^3} + 3\vb{T}\vb{w}^2\delta\vb{w} + \delta\vb{T}\vb{w}^3 &= \cancel{\lambda\vb{w}} + \delta(\lambda\vb{w}) \\
        \cancel{3 \lambda\vb{w}^2 \delta\vb{w}} + \delta\vb{T}\vb{w}^3 &= \delta(\lambda\vb{w})
    \end{aligned}
\end{equation}
We have used, respectively, the eigenvector definition,
the contraction property $\vb{T}\vb{w}^2 = \lambda\vb{w}^2$
and the orthogonality property $\vb{w}\cdot\delta\vb{w} = 0$.
This gives the desired sensitivity expression.

\end{appendices}

\newpage

\bibliographystyle{plain}
\bibliography{main}

\end{document}